# Strain Anisotropy Driven Spontaneous Formation of Nanoscrolls from Two-Dimensional Janus Layers

Mohammed Sayyad[1], Ying Qin[1], Jan Kopaczek[1,2], Adway Gupta[3], Naim Patoary[4], Shantanu Sinha[1], Emmie Benard[1], Austin Davis[1], Kentaro Yumigeta[3], Cheng-Lun Wu[1], Han Li[3], Shize Yang[1], Ivan Sanchez Esqueda[4], Arunima Singh[3], and Sefaattin Tongay[*1]

[1] Materials Science and Engineering, School for Engineering of Matter, Transport and Energy, Arizona State University, Tempe, AZ 85287, USA
[2] Department of Semiconductor Materials Engineering, Faculty of Fundamental Problems of Technology, Wroclaw University of Science and Technology, Wybrzeże Stanisława Wyspiańskiego 27, 50-370 Wrocław, Poland
[3] Department of Physics, Arizona State University, Tempe, AZ 85287-1504, USA
[4] Electrical, Computer and Energy Engineering, Arizona State University, Tempe, AZ, 85281, USA

Corresponding Author: sefaattin.tongay@asu.edu

## Abstract

Two-dimensional Janus transition metal dichalcogenides (TMDs) have attracted attention due to their emergent properties arising from broken mirror symmetry and self-driven polarisation fields. While it has been proposed that their vdW superlattices hold the key to achieving superior properties in piezoelectricity and photovoltiacs, available synthesis has ultimately limited their realisation. Here, we report the first packed vdW nanoscrolls made from Janus TMDs through a simple one-drop solution technique. Our results, including *ab-initio* simulations, show that the Bohr radius difference between the top sulphur and the bottom selenium atoms within Janus $M_{Se}^{S}$ (M=Mo, W) results in a permanent compressive surface strain that acts as a nanoscroll formation catalyst after small liquid interaction. Unlike classical 2D layers, the surface strain in Janus TMDs can be engineered from compressive to tensile by placing larger Bohr radius atoms on top ($M_{S}^{Se}$) to yield inverted C scrolls. Detailed microscopy studies offer the first insights into their morphology and readily formed Moiré lattices. In contrast, spectroscopy and FETs studies establish their excitonic and device properties and highlight significant differences compared to 2D flat Janus TMDs. These results introduce the first polar Janus TMD nanoscrolls and introduce inherent strain-driven scrolling dynamics as a catalyst to create superlattices.

## Introduction

Two-dimensional Janus transition metal dichalcogenides (TMDs) layers are an emergent class of atomically thin materials derived from their parent 2D classical TMDs systems[1-4]. Unlike the classical two-dimensional TMDs (MoSe$_2$, WSe$_2$, etc.) with the same chalcogen termination along either side of the basal metal plane, 2D Janus TMD layers contain different chalcogen atoms on either side. The chalcogen atoms in these 2D Janus TMDs are not intermixed (alloyed); instead, two different chalcogen atoms are present on opposite sides. To prevent confusion from 2D TMD alloys (such as MoS$_{2x}$Se$_{2(1-x)}$), Janus materials are described with $M_{Y}^{X}$ notation where X and Y (X≠Y=S, Se, or Te) are the top (X) and bottom (Y) chalcogen layers. Their unique crystallographic properties [5,6], broken mirror symmetry [3], and colossal out-of-plane polarisation field [2,3] placed 2D Janus materials at the frontiers of nanomaterials fields. Predictions to date have shown that many exciting quantum properties and functionalities can be realised; Examples include an immense E-field-induced DMI [7] and Skyrmion formation [8], enormous Rashba effect, a considerable Berry curvature value [2,9-12], E-field driven catalysis [13-20], substantial piezoelectricity [5,6], bipolar excitons [21], and spin-manipulation without the need for a magnetic field [22]. Furthermore, seminal studies have shown that these Janus layers can be experimentally realised using highly specialised techniques in 2D planar form [23-29].

While building materials in the 2D landscape offer clear advantages, more complex geometries, including superlattices, nanotubes, or nanoscrolls [30,31] create new directions and opportunities in fluids, energy storage, photovoltaics, sensors, and photonics. These higher-order superlattices allow for exotic structures that are not feasible using conventional manufacturing techniques and can enable new functionalities due to enhanced layer-to-layer interactions [32-34]. Indeed, recent studies have shown that flat 2D classical layers can be rolled to form nanoscrolls through hydrothermal [32], plasma [35,36], or solution-enhanced intercalation methods [32]. The latter technique, for example, is scrolling flat CVD-grown 2D TMDs to nanoscrolls solely by releasing (thermally introduced) substrate/material interface strain [32].

Here, our results mark the first experimental realisation of nanoscrolls from 2D Janus layers and establish their unusual scrolling dynamics driven by their inherent mechanical properties. Scrolling 2D Janus layers presents an entirely different and peculiar case owing to large mechanical anisotropy on their different surfaces and finite polarisation fields. The surface with disproportionately large Bohr radius chalcogen atoms retains tensile strain, while the opposite side with smaller Bohr radius chalcogens experiences compressive strain[37]. This strain polarisation [37] acts as a catalyst to readily form nanoscrolls with a minor agitation, in stark contrast to 2D classical TMDs that require an artificial substrate/material strain. Interestingly, 2D Janus layers with top compressive strain ($MoS_{Se}^{S}$) form well-defined nanoscrolls while the top surface with tensile strain ($MoS_{S}^{Se}$) forces 2D layers to tuck under with different topologies. *Ab-initio* molecular dynamic simulations and first principles DFT calculations offer first insights into the preferential direction of rolling and mechanics of the scrolling process. Detailed microscopy and spectroscopy measurement studies further demonstrate the unique Moiré lattice and establish the excitonic properties of 2D Janus nanoscrolls. Field effect transistors fabricated from these nanoscrolls show stable device operation. Overall results introduce nanoscrolls made from 2D Janus layers and a new kind of nanoscroll formation driven by the inherent mechanical properties of 2D Janus sheets.

**Results**

2D Janus TMDs were created from their classical TMD parent (Mo/W) by replacing the top layer selenium atoms with sulphur using an in-situ selective epitaxial process [24,25,38]. The classical 2D TMD monolayers were synthesised by the chemical vapour technique (CVD) established in the literature [39,40] (see **Fig.1a**) or by mechanical exfoliation from flux-grown layered TMD crystals by using a polymer-assisted exfoliation as shown in **Fig. S1a**. [41,42]. Starting with the classical 2D MoSe$_2$ and WSe$_2$ monolayers, the top layer selenium atoms are removed in the presence of reactive H$_2$ plasma, and open bonding sites are replaced by sulphur atoms assisted by H$_2$S gas without breaking the vacuum, as reported in earlier studies[24,25,38]. The energetics of the reaction relied on finding suitable reactive gas species and ionisation potential (plasma energy) of radical species (H$_2$ and H$_2$S) to promote the removal of M-Se bonds while curing these sites with stronger M-S bonds. The process was monitored by in-situ Raman and PL spectrometers integrated into our custom-made glass chamber **(Fig.1b)** to ensure Janus layers were synthesised and exhibited high photonic and PL characteristics. A typical in-situ process produced the most prominent Raman peak for $W_{S}^{Se}$ at 284 cm$^{-1}$ and for $MoS_{S}^{Se}$ at 292 cm$^{-1}$ for A$_1$ vibrational modes, as shown in **Fig. 1c,** and is consistent with the published values [9,10,23]. Similarly, PL emission lines were located at 1.80 eV ($W_S^{Se}$) and 1.71 eV ($Mo_S^{Se}$) **(Fig.1f)** compared to 2D classical WSe$_2$ and MoSe$_2$ neutral exciton emissions at 1.64 eV and 1.55 eV, respectively [4,23-25,38].

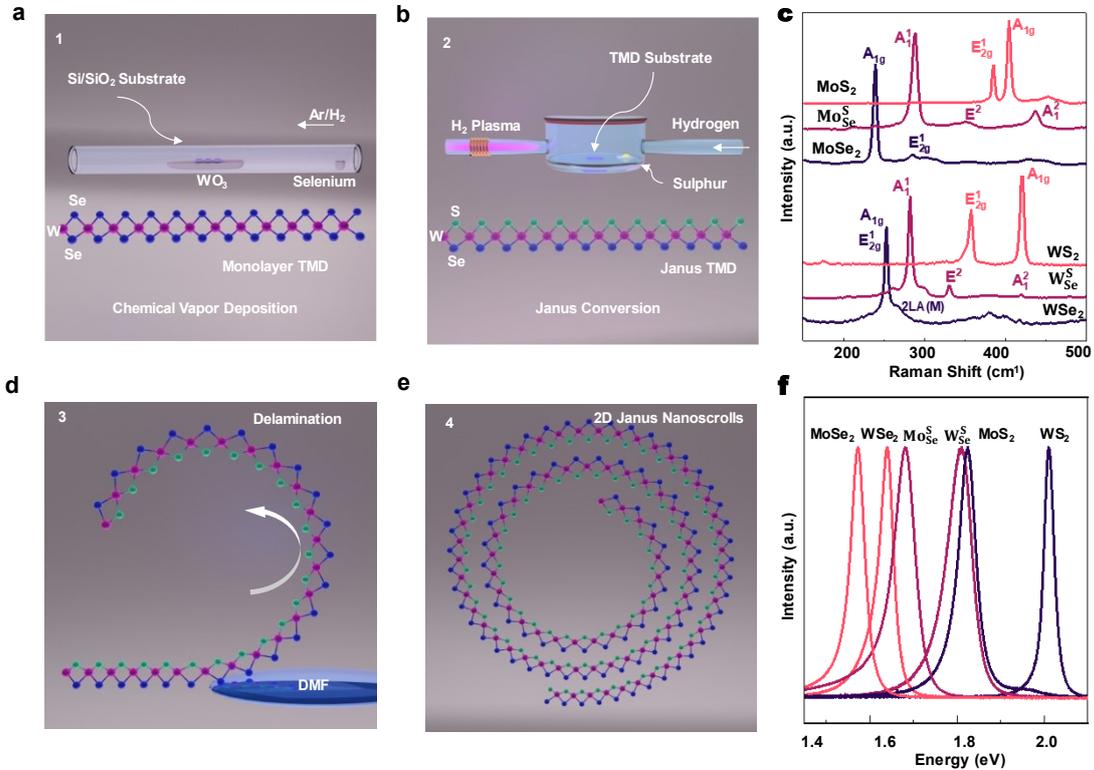

**Figure 1:** Schematic representation of Janus Nanoscrolls formation. **a.** Atmospheric Pressure Chemical Vapor Deposition of monolayer Transition Metal Dichalcogenide – WSe$_2$ **b.** Selective Epitaxy and Atomic Replacement Process carried out in prototype quartz chamber to form the Janus monolayers $W_{Se}^S$. **c.** Raman spectra and characteristic modes Classical TMDs and their Janus counterparts. **d**. (DMF) induces decoupling of the Janus monolayer from the Si/SiO$_2$ substrate and causes the formation of Janus Nanoscrolls via assisted strain release under the influence of 2,5 -Dimethylfuran (DMF) **e.** Schematic representation of quasi one-dimensional Janus Nanoscroll. **f.** Photoluminescence spectra of monolayer Transition Metal Dichalcogenides and their Janus counterparts.

Once these 2D Janus TMDs (originating from exfoliated or CVD-grown TMDs) are exposed to a small amount (50μL) of 2-5 Dimethyl Furan (DMF), they initiate the curling behaviour as schematically shown in **Fig.1d-1e**. The monolayer decouples from the substrate and readily rolls to form several small domains of nanoscrolls **(Fig. 2a and S1.c)**

As shown in **Fig. 2a and 2e**, exposed 2D Janus TMDs instantaneously decouple from the substrate and readily roll into several small domains of nanoscrolls **(Fig 2a and Fig. S1b** inset shows the $W_{Se}^S$ and $Mo_{Se}^S$ Janus monolayer before DMF treatment, respectively**.** The atomic force microscopy (AFM) images collected from different regions at magnifications **(Fig.2b, 2c and 2d)** clearly show the formation of quasi-1D nanoscrolls of $W_{Se}^S$ and $Mo_{Se}^S$ **(see Fig. S1. d)** from exfoliated TMDs, similarly as a comparison, **Fig S6** shows the AFM topography of $W_{Se}^S$ Janus nanoscroll created from the CVD-grown WSe$_2$ shown in the optical image of **Fig. 2e.** Based on our AFM characterisation on these exfoliated nanoscrolls, the average outer diameter ranges around 15-45 nm **(Fig. 2f )** and lengths of about 200 nm – 1μm **(Fig. 2g),** Interestingly enough, the nanoscrolls from exfoliated sources do not show any preferential orientation with respect the bulk crystal and the substrate and are distributed randomly with most of the population oriented between 0° and 45 ° with respect to the horizontal scan direction (left to right) as shown in **Fig. 2h** . The equilibrium outer diameter values correspond to ~7 to 20 rolls under the assumption that layer-to-layer distance is close to the equilibrium interlayer distance (3Å) and expected thickness of a monolayer Janus sample (7Å), We note, however, that the actual number of rolls can significantly vary due to variation in the actual equilibrium distance during the chemical processing. Similar scrolling behaviour was also observed on 2D Janus TMDs from CVD-grown classical 2D TMDs, as shown in **Fig. 2e** (*2e inset shows the Janus monolayer from CVD-grown WSe₂ before DMF treatment* ).

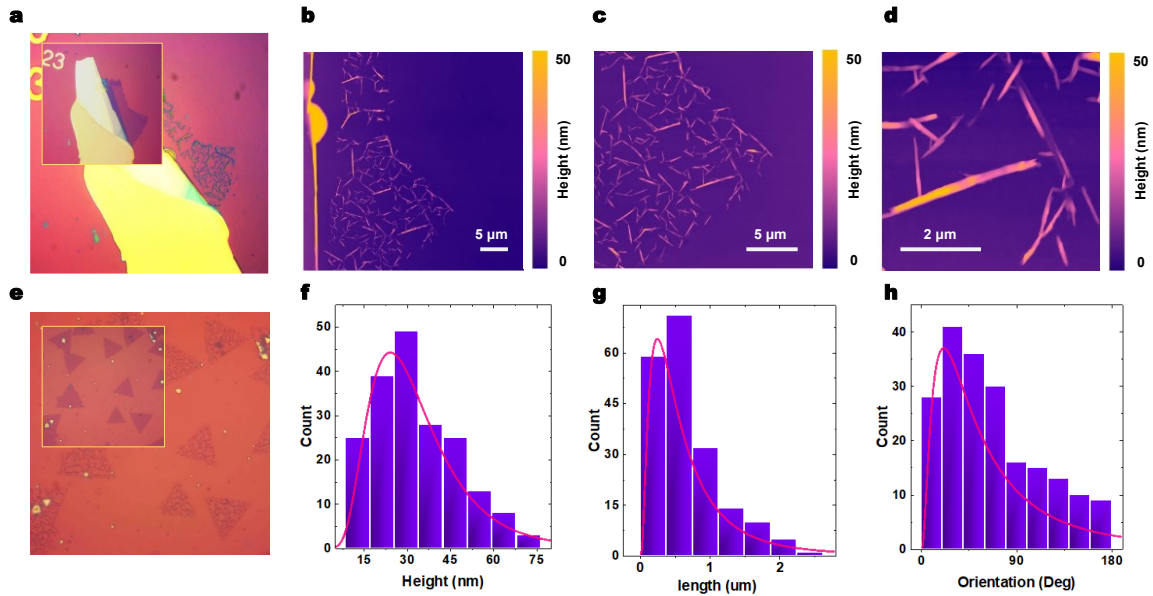

**Figure 2: a.** Monolayer Janus TMDs (from exfoliated TMDs) before and after treatment with 2-5 dimethylfuran **b-d.** Atomic Force Microscopy images collected from 2D Janus nanoscrolls. **e.** Before and after scrolling behaviour of 2D Janus $W_{Se}^{S}$ was created from CVD-grown WSe₂ monolayers. Representative statistical distribution of the **f.** height profile, **g.** length, and **h.** orientation of 2D Janus $W_{Se}^{S}$ Nanoscrolls with respect to the horizontal scanning direction of the AFM probe (left to right) are depicted in Fig.2b.

While some limited scrolling behaviour was also observed in classical TMDs [32,34,36], the observed behaviour on 2D Janus TMDs has stark differences; *Firstly*, exposing 2D exfoliated classical TMDs to the same liquid media does not result in any nanoscroll formation. This is also consistent with earlier studies on exfoliated classical TMDs wherein no roll-up response was found after exposure to a wide variety of liquid media [32,35,36]. The only examples of classical TMDs nanoscrolls originate from monolayer samples produced by hydrothermal[35,36] or on CVD-grown 2D classical TMDs sheets.[32] It is vital to note that unlike the previously established studies, where the driving force for scrolling in CVD-grown TMD was attributed to the "extrinsic" strain that is developed due to the difference in the thermal expansion coefficients of the monolayer and the substrate, the Same TMDs exfoliated from a bulk crystal will not undergo scrolling. In stark contrast, Janus-based TMDs undergo scrolling regardless of the synthesis method, suggesting that the rolling behaviour of these materials is a (1) intrinsic property, (2) dependent on the structural anisotropy and (3) independent of the substrate interaction. The *second* and most striking difference is observed when atoms are arranged in an opposite configuration to form $W_S^{Se}$ or $Mo_{Se}^{S}$ layers (instead of $W_{Se}^{S}$ or $Mo_{Se}^{S}$) by replacing the top sulphur layer of exfoliated 2D classical WS$_2$ / MoS$_2$ with selenium in the presence of H$_2$ and H$_2$Se gases (see methods). Once 2D $W_{S}^{Se}$ Janus monolayers are exposed to the same solution (50 μL of DMF), they tend to form inverted C-shaped scrolls by curling monolayers in-the-plane (or towards the substrate **(Fig. 3a)**. This behaviour is opposite to nanoscroll formation, as observed in **Fig. 3b**. for $W_{Se}^{S}$ wherein the monolayer sheets tend to scroll out-of-the-plane (away from the substrate. This suggests that in both cases, i.e., $W_{Se}^{S}$ and $W_{S}^{Se}$ the rolling direction is always favoured towards the atom with the smaller Bohr radius (i.e.) sulphur side. **Fig.3e and 3f**. show the schematic representation of these C-scrolls and Nanoscrolls for the Selenium atom facing away and towards the substrate, respectively. **Fig. S2.a and S2.e** show similar curling dynamics for Molybdenum-based TMDs starting from MoS$_2$ and MoSe$_2$, respectively (**Fig. S2. and S2.e** *Insets* show Janus $Mo_{S}^{Se}$ created from monolayer MoS$_2$ and Janus $Mo_{Se}^{S}$ created from monolayer MoSe$_2$ before the DMF treatment respectively).

Scanning Electron Microscopy **(Fig. S4 a-d)** and Atomic Force Microscopy measurements also confirm the behaviour of the curling dynamics in Molybdenum based systems (See **Fig. S2. b-d** (buckled) and **Fig. S2. f-h** (scrolled)) and further suggest the findings from Figure 3 panel to be universal across all transition metal-based Janus TMDS synthesised to date. **Fig. S3.a** shows the phase-contrast AFM scan of 2D Janus $W_{Se}^{S}$ from WS$_2$ as well as WSe$_2$ as the starting parent TMD, respectively **(Fig. S3. b).** We attribute these different curling dynamics to sulphur ($W_{Se}^{S}$) and selenium ($W_{S}^{Se}$) top surfaces to compressive strain found on sulphur and tensile strain on selenium side, which will be referred to as strain-polarisation on 2D Janus TMDs as described below.

High Resolution - Transmission Electron Microscopy (HRTEM) also confirms the scrolling and buckling behaviour for Tungsten Based Janus TMDs. Fig. S5 Shows the scrolling differences based on the orientation of the smaller Bohr radius with respect to the substrate. For the case of the Janus monolayers with sulphur atoms towards the substrate side ($M_{S}^{Se}$), a buckled structure formed due to (1) Curling, or rolling, is always energetically favoured towards the atom with a smaller Bohr radius and (2) since the atom with the smaller Bohr radius, i.e. in this case, sulphur is towards the substrate, the interaction from the latter results in this buckled structure. When such a sample is decoupled from the substrate and transferred onto the TEM Grid, the structure relaxes and continues rolling till equilibrium is achieved (See Fig S5 a-c). Note: for both cases, scrolling happens towards the sulphur side. However, In the case of Janus Monolayers with a Sulphur atom away from the substrate ($M_{Se}^{S}$), the resultant DMF interaction causes the monolayer to scroll cleanly, forming the so-called Nanoscrolls. (See Fig. S5 d-f) When picked up from the substrate onto the TEM Grid, this sample maintains its morphology and undergoes no further relaxation.

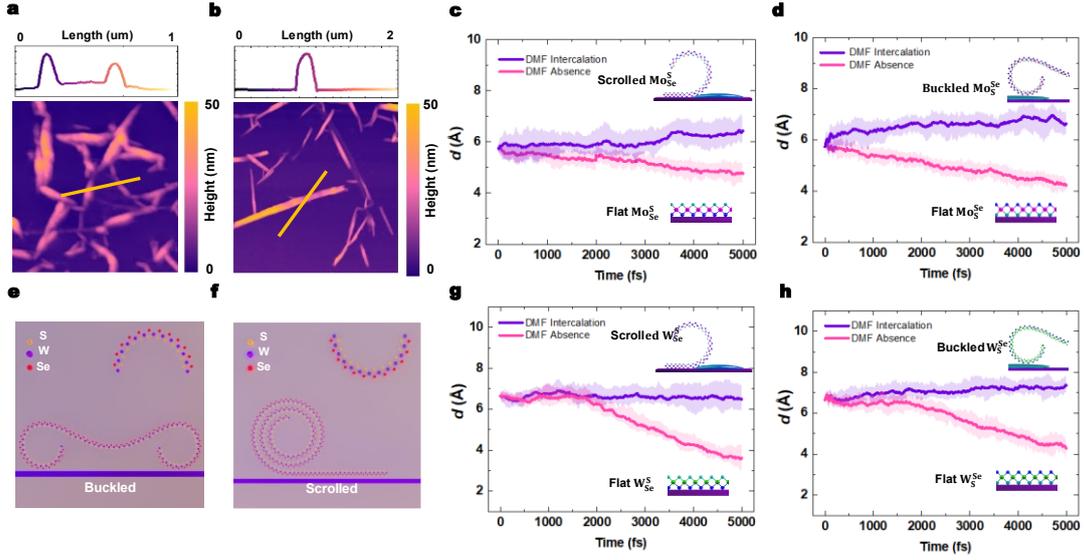

**Figure 3: a.** Atomic Force Microscopy on exfoliated Janus $W_S^{Se}$, with the sulphur plane resting on the substrate. **b.** Atomic Force Microscopy on exfoliated Janus $W_{Se}^{S}$, with the selenium plane resting on the substrate. *Ab-initio molecular dynamics simulated* **c.** z separation *(in Angstroms, Å)* of Janus $Mo_{Se}^{S}$ with the selenium plane on the substrate **d.** The *z*-separation of Janus $Mo_S^{Se}$ with the sulphur plane on the substrate. **e.** Schematic representation of a C – scrolled or buckled structure created from Janus $W_S^{Se}$. **f.** Schematic representation of a fully scrolled structure created from Janus $W_{Se}^{S}$. **g.** z separation of Janus $W_{Se}^{S}$, selenium plane on the substrate **h.** z separation of Janus $W_S^{Se}$, sulphur plane on the substrate. In **c, d-g, h,** the confidence interval, CI of the z-separation is plotted as shaded regions. Additionally, the structures in the inset in **c, d-g, and h** are representative schematics of scrolled/buckled nanosheets.

**Role of surface strain on scrolling:** In general, the scrolling dynamics can be described by two competing energy contributions that dictate the initiation and nanoscroll formation of Janus layers; (1) elastic energy that increases with the curvature of the sheet, effectively opposing bending, and (2) the reduction in the total free energy by stacking layers onto each other by weak vdW interactions [30]. In free self-standing form, the first elastic term increases the system's total energy by bending due to the restoring force of the mechanical strain. However, 2D CVD-grown classical TMDs retain strain at the material/substrate generated during the CVD cooling process (due to different thermal expansion coefficients of the constituent layers.) Small liquid exposure and concurrent lifting at edges by the liquid intercalation can initiate the rolling process and initially reduces the elastic term by releasing this extrinsic strain. [30] As such, the interface strain effect only applies to CVD-grown layers and excludes exfoliated ones which is the primary reason for no nanoscroll formation in exfoliated TMDs [32].

Here, comprehensive van der Waals (vdW) corrected DFT simulations were performed to offer theoretical insights into the preferential direction towards the S-top surface side (in $Mo_{Se}^{S}$ and $W_{Se}^{S}$). To compare the relative stability of $M_Y^X$ (M= Mo/W and X= S, Y=Se) sheets subjected to a positive curvature (curling towards the Y-side) *vs* negative curvature (curling towards the X-side), we compute the areal bending energy density, defined in *Equation 1* as:

$$\Delta E_{bend} = -\frac{E_{flat} - E_{curve}}{A_{flat}} \qquad eq.1$$

where $E_{flat}$ is the total energy of a flat Janus sheet $M_Y^X$, $E_{curve}$ is the energy of the curved sheets and $A_{flat}$ is the surface area of the perfectly flat 2D sheet. Additionally, to establish the role of edges in relative stability, we consider nanoribbons with both armchair and zigzag edges. To void interactions from the periodic cells, these Janus sheets are simulated in a slab geometry with more than 20 Å vacuum in two of the three orthogonal directions of the slab (see **Fig. S7 a-f**). **Table 1** reveals that the areal bending energy density of the Janus sheets is lower when the sheets curve towards the S-side compared to the Se-side, which is consistent with our experimental results. Moreover, these results remain consistent for varying curvatures and widths of the nanoribbon **(See Fig. S23, Table S2-S3)**

| Material | Edge | Curved towards | $\Delta E_{bend}$ (meV/Å$^2$) |
|---|---|---|---|
| $Mo_{Se}^S$ | Armchair | S-side | -1.407±0.003 |
| | | Se-side | 8.754±0.003 |
| | Zigzag | S-side | -1.281±1.146 |
| | | Se-side | 8.161±1.146 |
| $W_{Se}^S$ | Armchair | S-side | -2.076±0.003 |
| | | Se-side | 13.880±0.003 |
| | Zigzag | S-side | -1.394±1.014 |
| | | Se-side | 9.383±1.014 |

**Table 1:** Table shows the values of $\Delta E_{bend}$ for both armchair and zigzag edge nanoribbons of $Mo_{Se}^S$ and $W_{Se}^S$ for curvature towards the S compared to the Se face.

Further room-temperature ab-*initio* molecular dynamics (AIMD) simulations were performed on 2D Janus layers in the presence of substrate and DMF to understand the role of the substrate and the solvent in the spontaneous curling and scrolling of the Janus sheets. **Fig S8a** shows the crystal structure for an $Mo_S^{Se}$ nanosheet placed on SiO$_2$ with intercalated DMF, and similar simulations were performed using $Mo_{Se}^S$, $W_{Se}^S$ and $W_S^{Se}$ (not shown). Additional simulations were performed without DMF molecules at the Janus/substrate interfaces to discern the role of DMF. Fig. 3c, 3d, 3g and 3h depict the *z*-separation between the 2D Janus and the substrate as a function of the simulated time. Where z-separation at each time step t, *d*(t) is defined in *equation. 2* as

$$d(t) = <\hat{z}(t)_{S/Se}> - <\hat{z}(t)_{Si\ surface}> \qquad eq.2$$

where $<\hat{z}(t)_{S/Se}>$ and $<\hat{z}(t)_{Si\ surface}>$ are the average z positions of the S (Se) layer facing the substrate and the surface Si atoms of the substrate (as shown in Fig. S8), respectively, both at the t$^{th}$ time step. **Fig.3c, 3d-**

**3g and 3h** show that when DMF is present at the interface, the 2D Janus layer decouples from the substrate (the *z*-separation increases) and scrolls. In contrast, when DMF molecules are absent, the 2D Janus layer tends to adhere to the substrate (z-separation decreases). During the 5 ps AIMD simulations, we observe that the DMF weakens the interaction between 2D Janus and the substrate, especially at those sites where DMF is present, by bonding with the substrate as evidenced by the overlapping DMF-substrate charge densities that indicate strong bonding (See Fig. S8). The theory results ultimately suggest that subjected DMF molecules first diffuse at the Janus/substrate interface, following DMF molecules bind to the substrate and effectively decouple the 2D Janus layers from the underlying substrate. Once 2D Janus layers are further decoupled from the substrate, the internal strain dynamics start to play a dominant role in a way that nanoscrolls are formed when the top surface has compressive strain (sulphur atoms on top). In addition, c-shaped scrolls are created when the top surface has tensile strain (selenium on top) **(Fig.3)**.

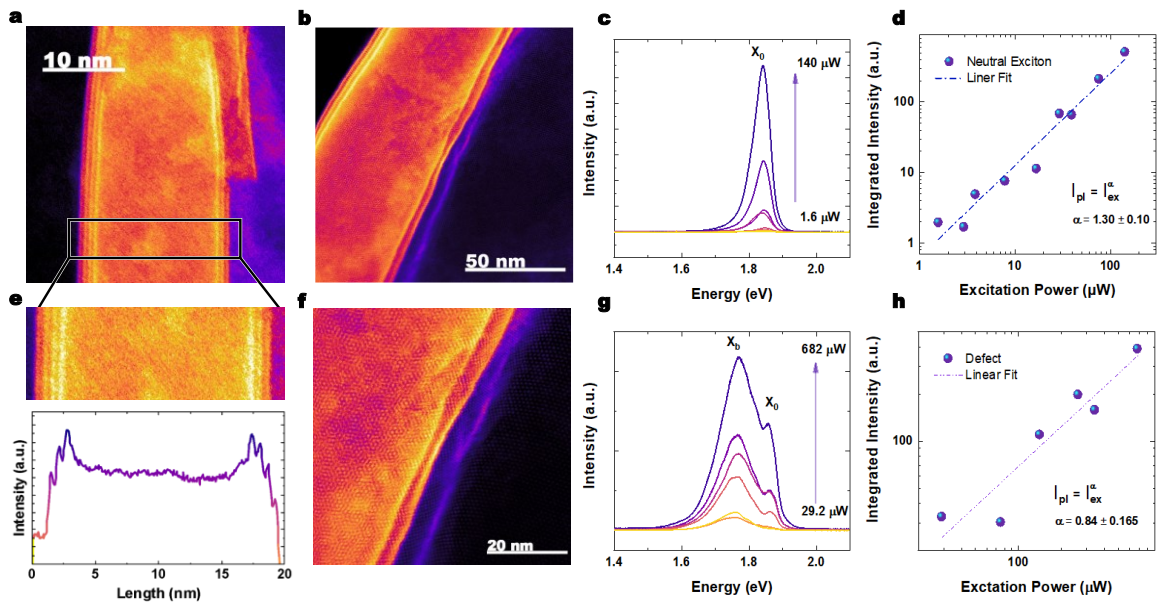

**Figure 4: a-b,** high angle annular dark-field (HAADF) scanning transmission electron (STEM) microscopy on 2D Janus $W^S_{Se}$ nanoscroll, **c.** Power-dependent photoluminescence spectroscopy on flat $W^S_{Se}$ sample, **d.** Integrated peak Intensity vs power from flat $W^S_{Se}$ **e.** HAADF STEM cross-section Janus $W^S_{Se}$ nanoscroll, bottom line intensity profile across the cross-section **f.** Formation of Moiré patterns on $W^S_{Se}$ nanoscrolls. **g.** Power-dependent photoluminescence spectroscopy on scrolled $W^S_{Se}$ sample. **h.** Integrated peak Intensity vs power from $W^S_{Se}$.

**Structural characteristics.** Scanning Transmission Electron Microscopy was used to characterise further the structure and morphology of the Janus nanoscrolls. As shown by the HAADF STEM images (Fig.4), the Janus nanoscrolls exhibited multi-walled and tubular structures with well-packed morphology that included weak (Fig. 4 a, b, and f) and strongly coupled (Fig. 4e) interlayer arrangements. Naturally, the nanoscroll outer diameter (based on height profile), length, and the number of rolls (layers) have been shown to vary (see **Fig. 2f, 2g and 2h),** depending on the initial Janus monolayer size as well as the scrolling dynamics when compared with nanoscrolls from traditional systems[43]. Here, closely packed Janus nanoscrolls exhibited an interlayer spacing of 6.32Å as determined by the STEM intensity mapping in **Fig. 4e** across the zoomed-in region from Fig. 4a, which is relatively close to the equilibrium interlayer spacing for vdW TMDs layers [44,45]. The typical Janus nanoscrolls can be described by the Archimedean spiral in *Equation 3*.

$$r = a\varphi + b \qquad\qquad eq.\ 3$$

where r is the radius, b is the initial core radii at $\varphi = 0$, $\varphi$ is the polar angle (ranging from 0 to $2\pi$), and a is the spiral constant given by $\frac{t}{2\pi}$ where the term *t* determines the change in the radius per one full turn. Using equation (3) and the data from STEM images, the nanoscroll in **Fig. 4e** can be described with b = 6.76 nm, t = 6.32 Å for N=3. As shown in **Fig.4f**, the roll-up process formed Moiré lattices (much like those recently observed in 2D $MoSe_2/WSe_2$ and $WSe_2/WS_2$ superlattices with exotic topological excitonic properties [46-48]). It is also noteworthy to mention that larger than anticipated interlayer spacing for some of the Janus nanoscrolls **(Fig.4a, 4b and 4f)** indirectly suggests that the energetic equilibrium is not reached between the layers, and the scrolling is primarily driven by the surface energy and strain polarisation as discussed in earlier sections.

**Optical performance.** The overall process was clean and did not cause visible damage to Janus layers, as confirmed by microscopy studies and temperature-dependent PL (See Supplementary Fig. S5, Fig. S13, and Fig. S14). **Fig. 4c and Fig. 4g** show power-dependent PL spectra collected from monolayer flat (un-scrolled) and nanoscroll $W^S_{Se}$ Janus samples at 77K, respectively. The typical PL spectra as a reference from $WSe_2$ before conversion is shown in Fig S15a. The three peaks are identified via power-dependent PL spectroscopy, as evidenced by the integrated peak intensity vs power graphs in **Fig. S16**. Following the SEAR conversion, the flat Janus $W^S_{Se}$ exhibits strong emission at 1.81 eV can be attributed to the neutral exciton emission [38] as confirmed by near-linear PL emission integrated area vs laser power density dependence **Fig. 4d** (also see supplementary Fig. S15b and Fig. S17 )[24,38]. Once scrolled, the primary PL emission appears at 1.76 eV with a small shoulder at 1.81 eV **(Fig. 4g)**. The PL spectra collected for increasing laser excitation show that the 1.81 eV peak becomes more pronounced at higher laser powers (See Supplementary Fig. S15c). We attribute the low energy (1.76 eV) peak to bound exciton emission due to its broad PL FWHM, sub-linear PL power dependence (See. Fig. 4h and SI Fig. S18), as well as similar emission line observed for defected flat Janus samples (See SI Fig. S12 - S13). Notably, even though the Janus nanoscrolls were rolled up to allow vdW stacking, the PL spectra retained 2D-like (electronically isolated) optical emission spectra, as shown in Fig. 4g, which is consistent with the STEM images in Fig.4a-d. Supplementary figure S15 a-c further compares excitonics states in monolayer $WSe_2$ with monolayer Flat Janus $W^S_{Se}$ and scrolled Janus $W^S_{Se}$. Typical emission spectra in monolayer $WSe_2$ at 77K consisted of sharp emissions at 1.73 eV and 1.70 eV attributed to the neutral exciton and Trion emission, respectively, as evident from linear power dependence see Fig. S16 b-c [38]. The emission spectra of the monolayer also exhibit a shoulder at 1.65eV that is attributed to bound exciton emission, as evident in the sub-linear power dependence (Fig. S16a).

As a quality measure, Raman full-width-at-half-maximum (FWHM) changed from 5.05775 to 5.36098 cm$^{-1}$, which is minuscule compared to standard 2D layer processing techniques[24,25,49,50]. Raman measurements also show that vibrational modes are blue-shifted at room temperature and 77 K, which can suggest stronger interfacial coupling (See Fig. S19-S20). From the perspective of strain, the 2D Janus nanoscroll show anomalous behaviour by blue shifting under a combination of tensile and compressive strain within layers[25].

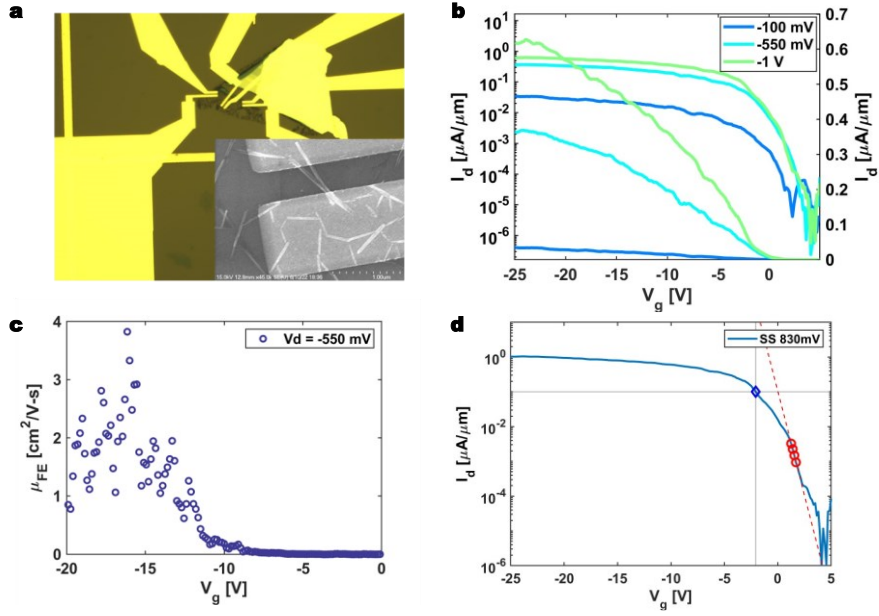

**Figure 5 a.** Optical image of backdated Janus $W_{Se}^S$ FET with a channel length of 370 nm. **b**. Transfer characteristics (Id-Vg) plotted in both log and a linear scale for SWSe scroll FET with Pd/Au contact as a function of drain voltage at room temperature **c.** extracted Field effect mobility **d.** extracted subthreshold slope

**FET Characteristics and performance.** Here, electrically active $W_{Se}^S$ layers within the nanoscroll naturally offer an added benefit by providing environmental inertness. The PDMS-aided transfer was used to transfer $W_{Se}^S$ nanoscrolls onto a pre-patterned 300 nm thick SiO$_2$/Si substrate. **Fig.5a** shows SEM micrograph and Pd/Au sources, as well as drain contacts, that were fabricated using E-Beam lithography with PMMA as a mask on top of the nanoscrolls Examples of $W_{Se}^S$ nanoscroll FET properties such as channel conduction properties with gate control (Id-Vg), Field effect mobility, and extracted subthreshold swing have been reported **(in Fig. 5b, 5c, and 5d).** In addition, the device exhibited an excellent electrical response, and the I-V behaviour was consistent with traditional FETs models[51]. Furthermore, the device showed consistent FET response under different gaseous environments owing to the much-protected (scrolled) structure during the annealing and vacuum cycles. While starting WSe$_2$ monolayers exhibited standard p and n-type conduction, the Janus nanoscrolls displayed only *p*-type conduction which is not fully understood at this time and merit future investigations. The overall field effect mobility was calculated using equation 4 [52];

$$\mu = \frac{d(\frac{Id}{Vd}\frac{L}{W})}{dVg} \cdot \frac{1}{Cg} \qquad \text{eq. 4}$$

Here, L and W are the channel length and width, respectively, µ is the field effect mobility, C$_g$ is gate capacitance per unit area (F/cm$^2$), and V$_g$ is the gate bias voltage. Extracted field effect mobility was in the 3-4 cm$^2$/V-s range **(Fig.5c),** with an on/off ratio > 10$^3$ for multi-walled nanoscrolls and > 10$^5$ for fewer walls. The Janus nanoscrolls FETs mobility values are comparable to flat Janus $Mo_{Se}^S$ (5-10 cm$^2$/V-s)[24]. We note that lower mobility values are likely related to considerably significant vacancy defects formed in Janus TMDs due to harsh plasma processing/chemical treatment steps during lithography [24,38]. However, improvements in mobility may be achieved through channel and gate-stack engineering, including optimisation of the number of walls in the scrolls. The extracted subthreshold swing was 830 mv/dec **(See Fig. 5d).** While nascent, Janus FETs offer good FET characteristics with good environmental inertness. They are ideal future testbed systems to tune FET characteristics by polarisation-field direction engineering.

**Conclusion.** Overall findings demonstrated the first high-quality vdW nanoscrolls from 2D Janus layers using a simple one-drop solution process and highlighted striking differences in scrolling dynamics compared to 2D classical (non-Janus) layers. The results show that different scrolling dynamics arise from the unique two-faced atomic arrangement and surface strain polarisation found in 2D Janus TMDS. By engineering the atomic surface type and the surface strain polarisation, the current work created ways to realise unique morphologies, such as Archimedean nanoscrolls and inverted C-shaped scrolls. Microscopy studies established their morphology and demonstrated the first Moiré patterns of 2D Janus sheets.

In contrast, spectroscopy and FETs studies further highlighted the optical and electrical differences between Janus nanoscrolls and 2D flat Janus TMDS arising from scrolling. Overall, findings have provided simple ways to create vdW superlattices of 2D Janus TMDs and highlighted the importance of their mechanical properties in the scrolling process. Further understanding of their properties and the effects of Janus polarisation fields will be essential to realising the superior functionalities of these materials.

**Methods**

**Chemical Vapor Deposition and Sample Preparation:** Monolayer transition metal dichalcogenides ($MoSe_2$ and $WSe_2$) were synthesised directly on the $SiO_2$/Si substrate following well-established parameters in the literature (Fig. 1a). The substrates were thoroughly cleaned before the synthesis with a piranha etchant followed by sonication with double deionised water. The sonication is done twice to remove any etchant traces. The substrates are further sonicated with Ethanol and Isopropyl alcohol for another five minutes before an Oxygen plasma treatment for fifteen minutes.

Two-dimensional Molybdenum / Tungsten Diselenide ($XSe_2$ M=Mo, W) was synthesised in an Atmospheric Pressure Chemical Vapour Deposition (AP-CVD) system consisting of one inch (1") inner diameter (ID) quartz tube supported on a single zone Lindberg Blue/M tube furnace. The quartz tube is thoroughly cleaned with soap and DI water before the synthesis. Less than 1 mg Molybdenum (VI) Oxide (*Sigma-Aldrich, ACS reagent, ≥99.5%*), used as a metal precursor, was evenly distributed in a 7 ml Coors Tek combustion boat. Silicon substrates (polished side downwards) are kept on the boat. The substrates were cleaned in a sonicator using IPA followed by acetone for 15 minutes and then treated with Ar plasma to remove surface contaminants. 50 μM PTAS solution is added to one of the substrates to promote nucleation during the synthesis. Powdered Selenium (Sigma-Aldrich 100 mesh, 99.99% trace metals basis) is used as the chalcogen source and kept as an excess reagent at the upstream side with respect to the gas flow. The temperature distribution at this zone during the reaction is approximately 300 °C. A process gas mixture in the form of Ar/H is introduced into the reaction chamber from the upstream side. The furnace is ramped up to 790 °C in sixteen minutes, during which Ar is introduced at a flow rate of 46 SCCM. The furnace is held at this temperature for five minutes with the introduction of 4 SCCM H2. After the synthesis, the furnace is allowed to cool down naturally, and $H_2$ gas supply is cut off, and the Ar flow rate is increased to 200 SCCM. Synthesis of monolayer Tungsten Selenide is conducted similarly with the following exceptions. 50 Mg Tungsten (VI) Oxide (Sigma-Aldrich, ACS reagent, ≥99.5%)) is used as a metal precursor and is evenly distributed in another Coors Tek combustion boa. The system is flushed with Argon for fifteen minutes before introducing 80 SCCM carrier gas (Ar: H The furnace is heated up to 850 °C in seventeen minutes and kept at 850 °C for twenty minutes before natural cooling).

**Crystal growth and mechanical exfoliation:** A similar process was carried out for monolayer TMD samples created using tape exfoliation from bulk vdW crystals. Bulk crystals of $MoSe_2$ and $WSe_2$ were grown using chemical vapour transport [53-55] and then exfoliated onto silicon substrates with a 300 nm thick thermal oxide. The substrates were pre-cleaned and treated similarly to the CVD substrates. Exfoliated monolayer TMDs were prepared from CVT-grown bulk crystal over several successive exfoliation steps using a Nitto Blue tape after the sample was thinned to the desired thickness. The blue tape was eventually stamped on a thin layer of PDMS. This step is added before the final stamping to reduce the amount of tape residue on the final substrate and further thin down the sample. The PDMS stamp is pressed against an ultra-clean Silicon substrate with 300 nm thick oxide. The substrate is treated with 75 W oxygen plasma after the chemical treatment to remove traces of carbon contamination and improve the stamp's wettability. The final transfer is done within 15 minutes of the oxygen plasma treatment. The PDMS silicon substrate is kept inside a vacuum chamber for another ten minutes to remove air pockets and improve the substrate stamp contact. Finally, the stamp is peeled off slowly, and the samples are examined under an optical microscope to identify the regions of interest. These are then confirmed using atomic force microscopy and the absence of indirect emission in the photoluminescence-spectra

**In-situ Spectroscopy and Janus monolayer formation:** Synthesis of 2D Janus $Mo_{Se}^{S}$ and $W_{Se}^{S}$ is carried out in a custom-built glass reaction chamber. The chamber consists of two glass arms that serve as an inlet/outlet manifold. Ultrapure hydrogen is introduced through the inlet of the chamber. The pressure within the chamber is regulated and monitored by a pressure gauge and capacitance manometer connected at the inlet and outlet manifolds. CVD-grown and mechanically exfoliated TMD monolayers serve as a reaction precursor and are kept at the centre of the synthesis chamber. A secondary chalcogen source (sulphur) on a quartz boat is held upstream to the gas flow. The outlet manifold of the chamber is connected to an Edwards vacuum pump and kept at $-3.6 < 10^{-3}$ Torr during the entire synthesis. Before the synthesis, the chamber is devoid of air and observed under a vacuum for fifteen minutes. 20 SCCM hydrogen is introduced into the reaction chamber and stabilised for another ten minutes before the synthesis. A copper coil wound around the outlet manifold of the chamber is connected to a SEREN 101R RF power source and an impedance-matching network. Hydrogen plasma is ignited by supplying 5 W power to the copper coil. The entire synthesis is carried out under a Renishaw Optical Spectrophotometer with a 488nm laser focused using a long-distance objective. Real-time optical spectroscopic data (Raman and Photoluminescence) is collected in situ and aids in determining the optimum conversion of the transition metal dichalcogenide into Janus monolayers. Minor power and flow rate adjustments are made during the entire conversion process based on the data collected in situ during the measurement.

*Similarly, the conversion parameters were tuned for the Synthesis of 2D Janus $W_{S}^{Se}$ and $Mo_{S}^{Se}$ from their classical sulphur-based counterparts was also carried out in our custom-built SEAR Chamber. The sulphur-based Transition metal dichalcogenide is kept inside the chamber, which is pumped down for fifteen minutes to remove* *the traces of any residual oxygen and contaminant gas during the SEAR conversion reaction. The substrates are kept about 2 cm closer to the plasma source; this is done to ensure that the energy density of the plasma is high enough to overcome the bond strength of the sulphur-metal bond, which is different from those of their selenium counterparts. Additionally, the chalcogen source is changed to a selenium ingot created for selenium pellets (99.999% trace metal basis, Sigma Aldrich). Before the SEAR reaction, the chamber is also thoroughly cleaned and baked to remove any traces of residual sulphur from previous SEAR reactions. Optimum experimental conditions are empirically determined over successive runs by carefully tuning the power of the RF source and optimising the gas flow. For a typical reaction, higher power from the RF source (16W) was observed and is again readily explained by the bond strength of the metal sulphur bond. The hydrogen flow rate, however, is optimised at 20 SCCM to keep the concentration and ionisation of the radicals the same throughout the reaction. We, however, note that conversion from the sulphur-based TMD is generally defective based on the broadening of the PL Peak emission when compared to SEAR-converted Se-based TMDs. This has been attributed to the higher knock-on potential of the selenium radical, which is higher in mass with respect to sulphur. SI Figure S24 Shows the conversion PL and Raman spectra of Janus $Mo_S^{Se}$ and $W_S^{Se}$ after sear conversion from their sulphur-based counterparts, in all four cases, the reaction is terminated after the a1g Raman modes disappear in their classical TMD Counterparts.*

**Synthesis of 2D Janus Nanoscrolls:** Synthesis of 2D Janus Nanoscrolls is carried out by decoupling the monolayer from the substrate by treating it with 50 μL of 2-5 Dimethylfuran. The organosulfur compound allows the decoupling of the monolayer from its substrate, followed by instantaneous scroll formation due to the out-of-plane structural anisotropy of the Janus monolayer. Then, the substrate is allowed to dry naturally, followed by vacuum annealing in a tube furnace at 60°C to remove any traces of the organic solvent. Fig. S25. Shows the efficiency of scrolling for different solvents Fig. S25 a-b before and After Ethanol treatment and Fig S25 c-d before and after acetone treatment based on reducing relative polarity (DMF>Ethanol>Acetone)for the same experimental time duration as DMF. In our scope of experiments, DMF is chosen for its highest relative polarity and the lowest retention time for effective scrolling conversion. This is to ensure and minimise the number of mechanical defects (Tears / Cracks) that may be generated due to the extended duration of the interaction of the Janus Materials with the solvent.

**Low-Temperature Raman and Photoluminescence:** Low-Temperature Raman and Photoluminescence spectroscopy was carried out in a Janis ST-500 cryostat at 77k. The cryostat was positioned under a Renishaw In-Via Raman spectrometer with a 488 nm laser. Spectra were collected from the same position on the monolayer sample before and after the SEAR process and before and after scrolling. The temperature of the cryostat was controlled using a lakeshore 338 temperature controller and a LabVIEW program. The cryostat is connected to a Pfeffer turbo and pumped down to $10^{-6}$ Torr to remove moisture from the main chamber. Liquid nitrogen flows into the cryostat through a transfer line evacuated overnight to minimise thermal losses. The flow rate of $LN_2$ to the cold finger is controlled with ultrapure nitrogen gas used to pressurise the dewar. For the power-dependent spectral emission, the laser power is controlled using the grey filter equipped within the spectrometer and later calibrated using an external power meter for greater accuracy.

**Atomic force microscopy measurements**: The surface topography is characterised by an NT-MDT modular AFM. Bruker TESPA V2 probe with an 8 nm tip radius is used to record the sample height in a semi-contact mode and 512-line resolution. The images are plotted and processed using Gwyddion software.

**Scanning transmission electron microscopy and High-Resolution Transmission Electron Microscopy (STEM and sample preparation):** A region of interest is selected on a CVD-grown monolayer Janus TMD nanoscrolls, and a holey carbon Quantifoil is placed on top of the sample. One drop of IPA solution is poured over the sample to ensure sufficient contact. The grid substrate assembly is then spin-coated with PMMA to help act as additional support, followed by etching in 30 wt.% KOH. The transferred TMD monolayer is carefully fished out of the PMMA bath and rinsed three times with DI water to remove any traces of KOH. Next, the PMMA thin film is dissolved overnight in acetone, followed by another DI water rinse. The Scanning Transmission Electron Microscopy images were collected on a Nion Ultra STEM 100 microscope operated at 100 keV, with a convergence angle of 33 mrad, a beam current of around 10-30 pA., and an inner collection semi-angle of 50 mrad. The sample was baked at 160 °C for 10 hours before the experiment. The images are filtered using the Gaussian blur function in the Digital Micrograph microscopy suite.

**Fabrication of Janus Nanoscrolls FETs:** 90 nm Si/SiO$_2$ Substrate has been patterned using the EVG Aligner 620 photolithography with a chrome shadow mask. The substrate's back oxide has been entirely etched using a 790 RIE oxygen plasma chamber. The PDMS-aided wet transfer was used to transfer CVD-grown converted Janus $Mo^S_{Se}$ Scrolls onto a pre-patterned 90 nm thick SiO2/Si substrate. The substrate was then annealed at 200 °C under vacuum with Ar/H$_2$ forming gas flow. A thin mask of PMMA has been spin-coated at 4000 RPM on top of the substrate. The source and drain region were exposed with JEOL 6000FS Electron Beam Lithography. A 1:3 MIBK and IPA were used to develop the pattern onto the channel region. Next, 5nm Ti and 35nm gold were deposited using Lesker 3 Ebeam evaporation at a rate of 0.5 Å s$^{-1}$. Finally, acetone was used to lift the mask, excluding the pattern region with a channel length of 1.7 μm.

**Density functional theory and *ab-initio* molecular dynamics**:

We perform all our vdW-corrected DFT calculations using the PAW method implemented in the plane-wave code VASP [56-58]. The vdW interactions are modelled using the optB88 functional proposed by Klimes et.al. [59][All the calculations are performed with a *k*-grid density of 60 Å$^{-1}$ in all three directions. The total energy is converged within 10$^{-7}$ eV at every ionic relaxation step, and the forces on the atoms are converted to 5 meV/Å. A sufficiently high plane-wave cut-off energy of 600 eV was chosen. For the DFT calculations of the semi-infinite flat and curved $M^X_Y$ Janus nanoribbons, the input structures are generated for armchair and zigzag nanoribbons with a radius of curvature of 30 Å(See Fig. S6a-f) and then allowed to relax keeping the central layer of Mo atoms fixed to keep the curvature fixed. The S and Se atoms on either side are allowed to relax as usual. The conclusions are drawn from the curved and flat $M^X_Y$ Janus nanoribbons simulations do not change if the curvature was 50 Å or 100 Å (see Table S2-3, Fig. S20). The unit cell is periodic along one direction while there is a vacuum spacing of at least 20 Å along the two orthogonal directions to minimise the interaction between periodic images.

We use the Hetero2D package [60] to identify the heterostructures of the Janus materials and SiO$_2$ lattice planes, which have a low lattice mismatch and are symmetrical with the two Janus materials. Further details about the simulated configurations, with and without intercalated DMF, can be found in section S2.

All the *ab-initio* molecular dynamics calculations are performed at room temperature(300K) and using an NPT ensemble with a target pressure of 0 kB. The dynamics are calculated by the method proposed by Parrinello et al [61,62]. We simulate 5000-time steps, each of 1 fs, for a total simulation time of 5 ps to ensure all our aggregate parameters have converged; see Supplementary Fig, S9-S12.

**Acknowledgements:**

S.T acknowledges primary support from DOE-SC0020653 (materials synthesis), Applied Materials Inc., NSF CMMI 1825594 (NMR and TEM studies), NSF DMR-1955889 (magnetic measurements), NSF CMMI-1933214, NSF 1904716, NSF 1935994, NSF ECCS 2052527, DMR 2111812, and CMMI 2129412. J.K. acknowledges support within the Bekker program from the Polish National Agency for Academic Exchange. The authors acknowledge the San Diego Supercomputer Center under the NSF-XSEDE Award No. DMR150006 and the Research Computing at Arizona State University for providing HPC resources, the use of facilities within the Eyring Materials Center at Arizona State University supported in part by NNCI-ECCS-1542160. This research used resources of the National Energy Research Scientific Computing Center; a DOE Office of Science User Facility supported by the Office of Science of the U.S. Department of Energy under Contract No. DE-AC02-05CH11231. AG and AS acknowledge support by NSF DMR under grant # DMR-1906030

# Supplementary Information : Strain Anisotropy Driven Spontaneous Formation of Nanoscrolls from Two-Dimensional Janus Layers

Mohammed Sayyad[1], Ying Qin[1], Jan Kopaczek[1,2], Adway Gupta[3], Naim Patoary[4], Shantanu Sinha[1], Emmie Benard[1], Austin Davis[1], Kentaro Yumigeta[3], Cheng-Lun Wu[1], Han Li[3], Shize Yang[1], Ivan Sanchez Esqueda[4], Arunima Singh[3], and Sefaattin Tongay[*1]

[1] Materials Science and Engineering, School for Engineering of Matter, Transport and Energy, Arizona State University, Tempe, AZ 85287, USA
[2] Department of Semiconductor Materials Engineering, Faculty of Fundamental Problems of Technology, Wroclaw University of Science and Technology, Wybrzeże Stanisława Wyspiańskiego 27, 50-370 Wrocław, Poland
[3] Department of Physics, Arizona State University, Tempe, AZ 85287-1504, USA
[4] Electrical, Computer and Energy Engineering, Arizona State University, Tempe, AZ, 85281, USA

Corresponding Author: sefaattin.tongay@asu.edu

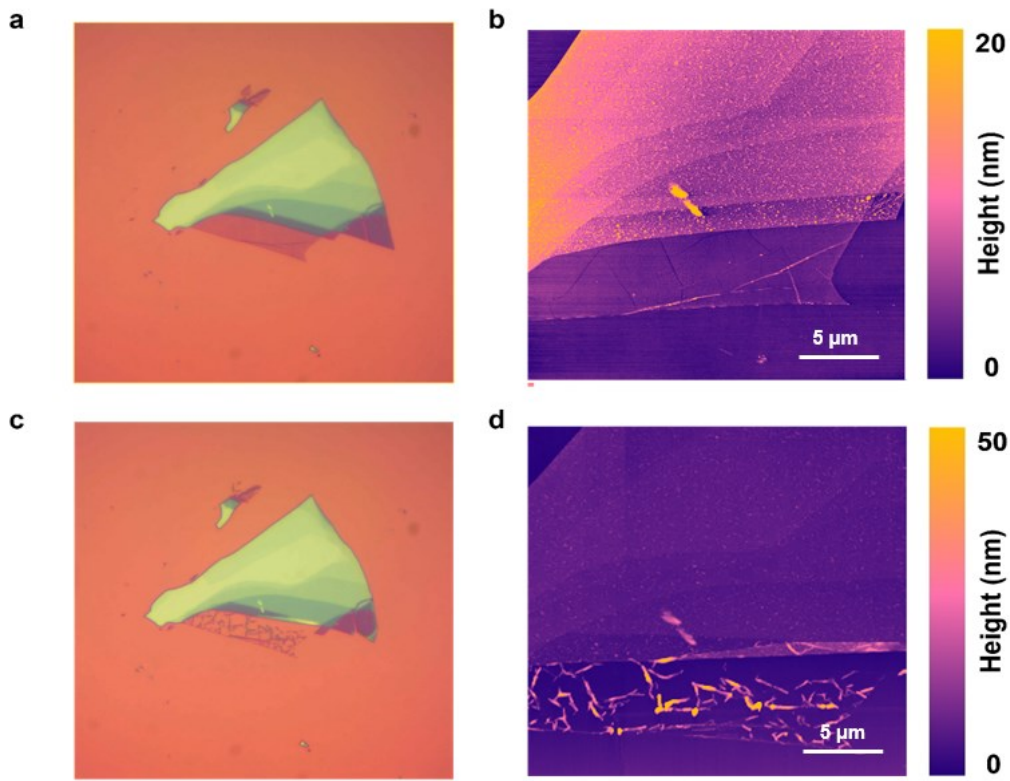

**Figure S1 a.** Monolayer Janus $Mo^S_{Se}$, obtained from Janus conversion of exfoliated flux grown MoSe$_2$ crystal before scrolling **b.** AFM scan of the flake in panel S1a **c.** Janus $Mo^S_{Se}$ after scrolling, showing the formation of Janus nanoscrolls after DMF treatment **d.** AFM scan of the nanoscrolls in panel c

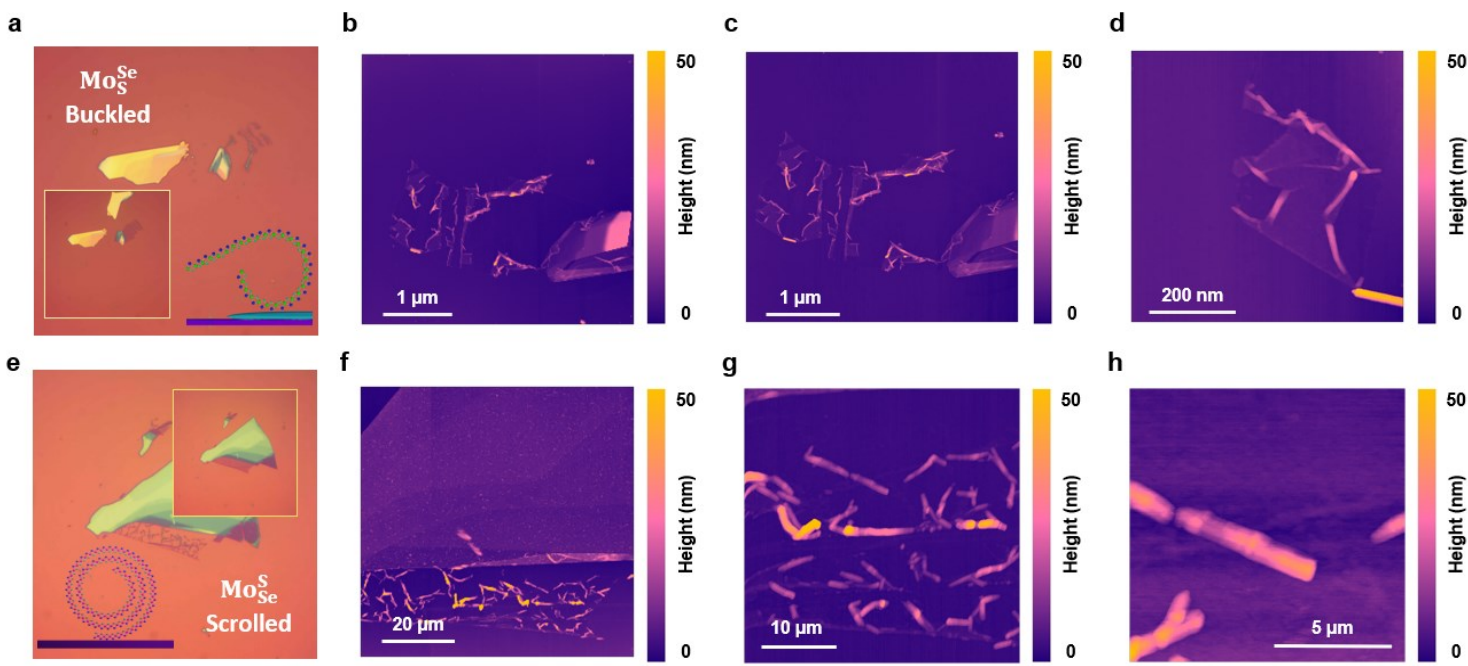

**Figure S2:** - **a.** Exfoliated Monolayer Janus $Mo_S^{Se}$ synthesised from MoS₂ bulk crystal, the inset shows the same monolayer before the DMF treatment **b-d.** AFM topography of Janus $Mo_S^{Se}$ Showing the buckled structure and formation of inverted C scrolls, **e.** Exfoliated Monolayer Janus $Mo_{Se}^{S}$ synthesised from MoSe₂ bulk crystal, the inset shows the same monolayer before the DMF treatment and nanoscroll formation **f-h.** AFM topography of Janus $Mo_{Se}^{S}$ showing the formation of isolated nanoscrolls.

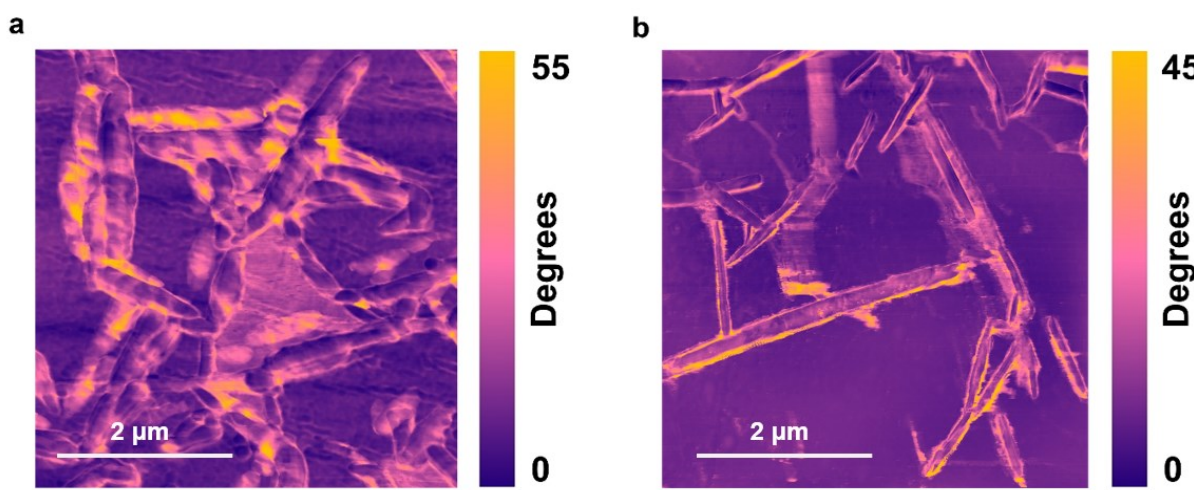

**Figure S3:** - **a.** Phase Contrast Force Microscopy on monolayer Janus $W_S^{Se}$ after DMF treatment, highlighting the formation of buckled structures b. Phase Contrast Force Microscopy on monolayer Janus $W_{Se}^{S}$ after DMF treatment, highlighting the formation of isolated nanoscrolls.

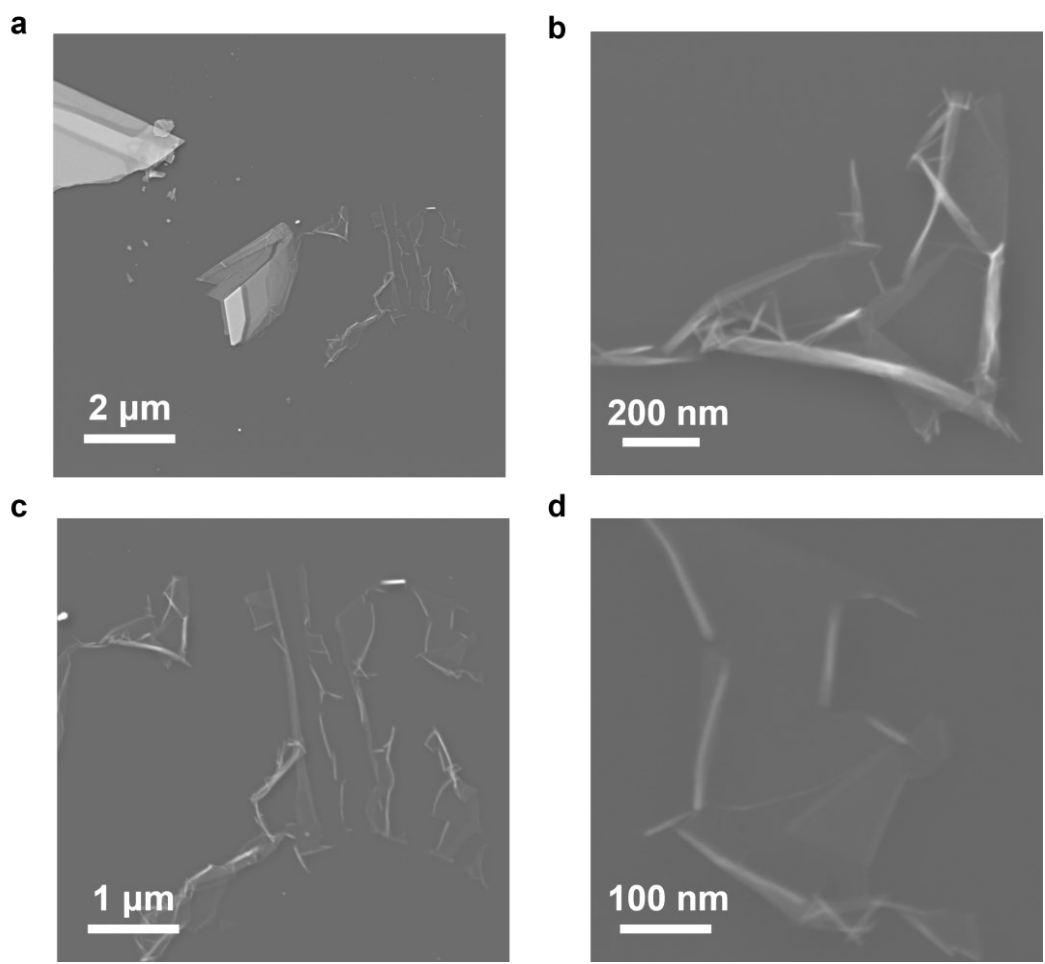

**Figure S4: - a-d** Scanning Electron Micrograph of Janus $Mo_S^{Se}$ a buckled structure created from monolayer MoS$_2$ as the starting parent TMD

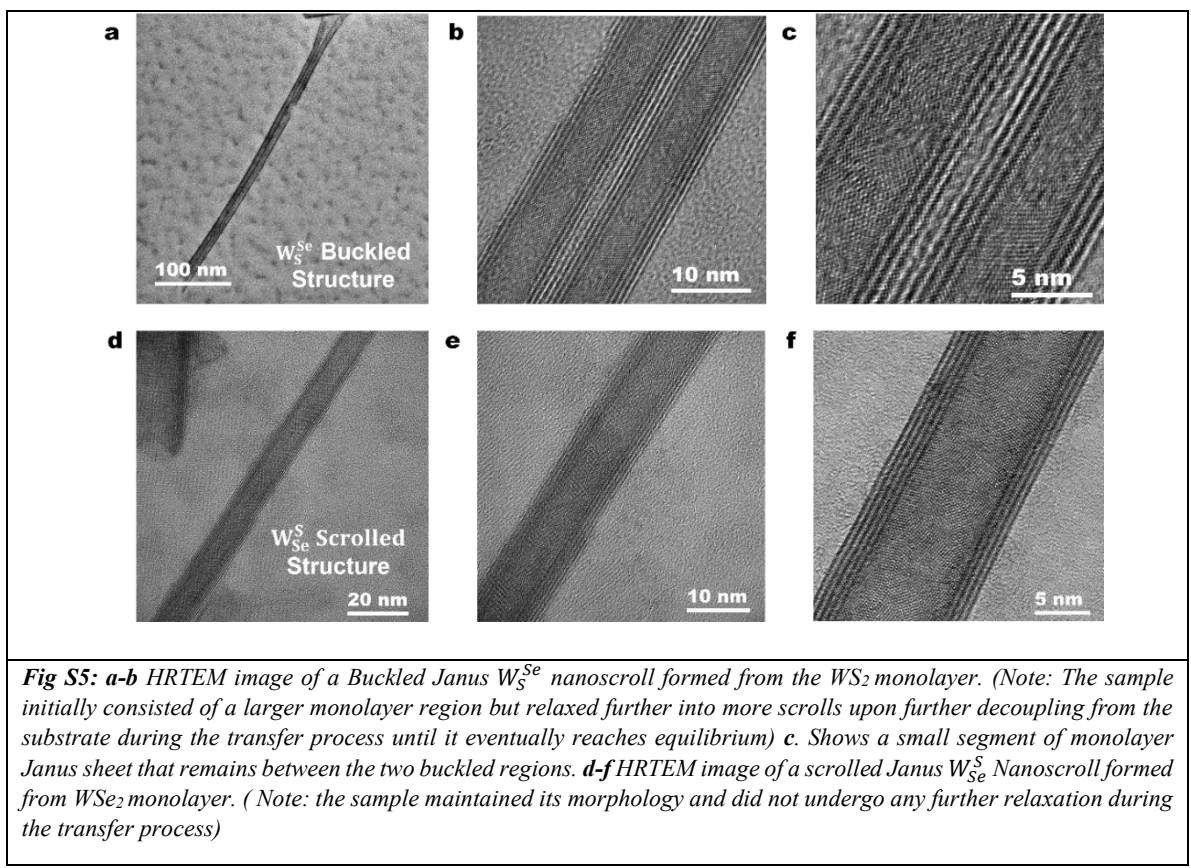

**Fig S5:** ***a-b*** *HRTEM image of a Buckled Janus $W_S^{Se}$ nanoscroll formed from the $WS_2$ monolayer. (Note: The sample initially consisted of a larger monolayer region but relaxed further into more scrolls upon further decoupling from the substrate during the transfer process until it eventually reaches equilibrium)* ***c****. Shows a small segment of monolayer Janus sheet that remains between the two buckled regions.* ***d-f*** *HRTEM image of a scrolled Janus $W_{Se}^{S}$ Nanoscroll formed from $WSe_2$ monolayer. ( Note: the sample maintained its morphology and did not undergo any further relaxation during the transfer process)*

Note: For the case of Janus Monolayers with a Sulfur atom away from the substrate, the resultant DMF causes the monolayer to scroll cleanly, forming the so-called Nanoscrolls. This sample maintains its morphology when picked up from the substrate onto the TEM Grid. For the case of Janus Monolayers with sulfur atoms towards the substrate side, a buckled structure is formed due to (1) Curling or rolling is always energetically favoured towards the atom with a smaller Bohr radius and (2) since the atom with the smaller Bohr radius, i.e., in this case, sulfur is towards the substrate, the interaction from the latter results in this buckled structure. When such a sample is decoupled from the substrate and transferred onto the TEM Grid, the structure relaxes and continues rolling till equilibrium is achieved. Note: for both cases, scrolling happens towards the sulfur side.

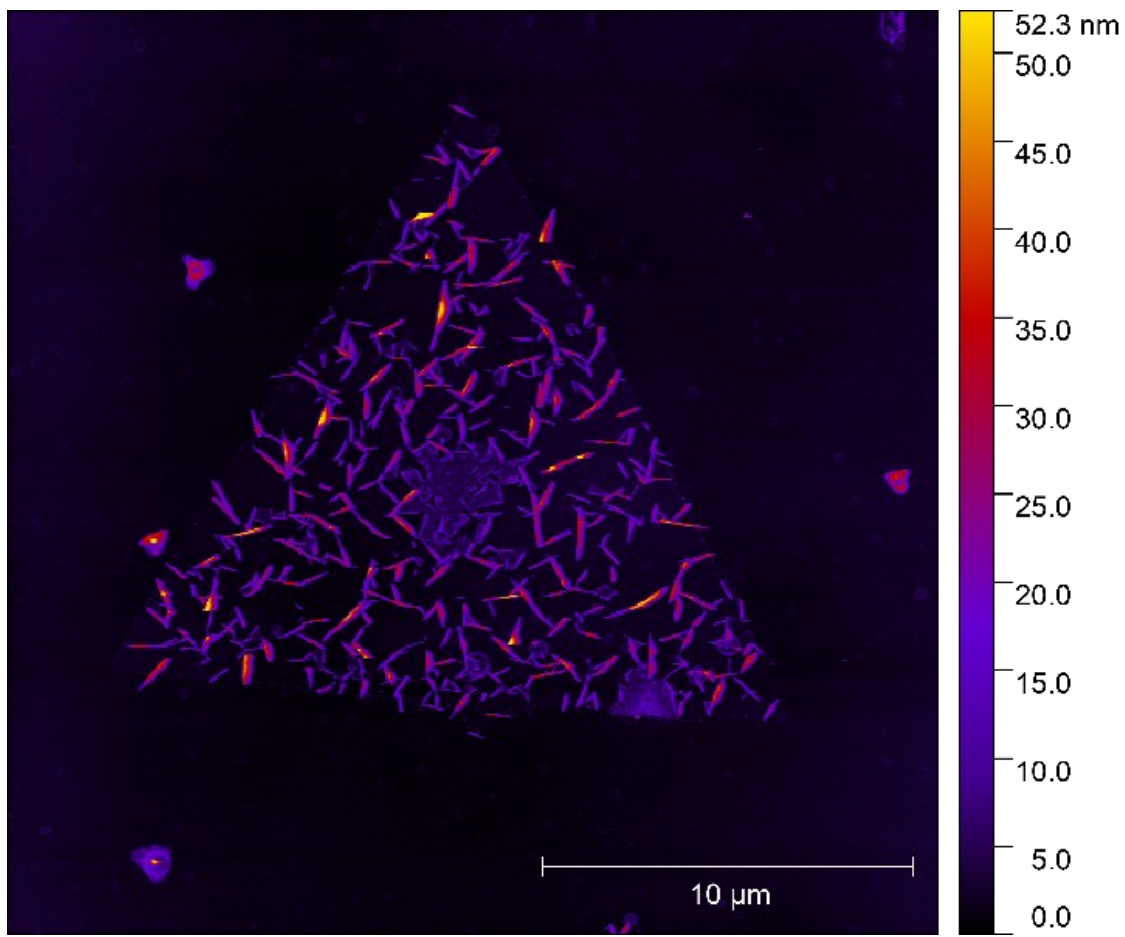

**Figure S6:** - an **AFM** scan of Janus $W_S^{Se}$ monolayer synthesised from monolayer WSe$_2$ grown from CVD

| Armchair Edges | Zigzag Edges |
|---|---|
| (a) Flat Mo$^S_{Se}$ nanosheet | (d) Flat Mo$^S_{Se}$ nanosheet |

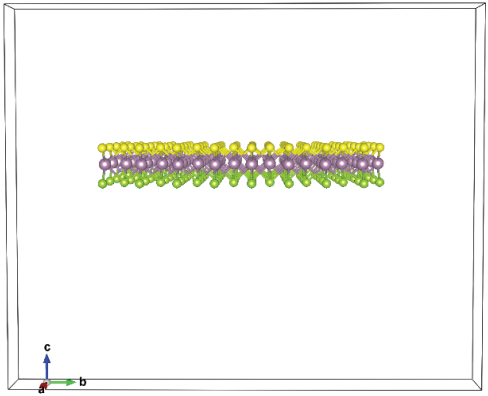 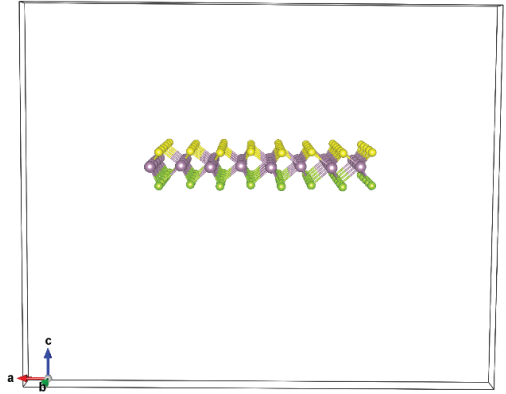

(b) Mo$^S_{Se}$ nanosheet curved towards S    (e) Mo$^S_{Se}$ nanosheet curved towards S

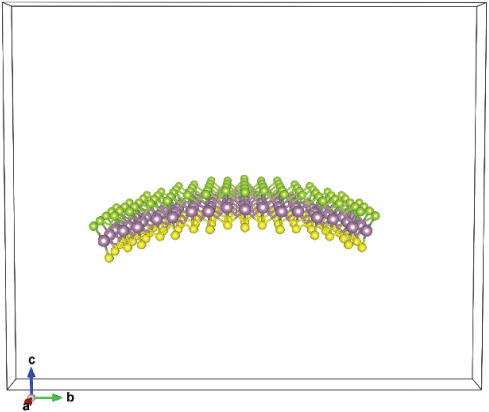 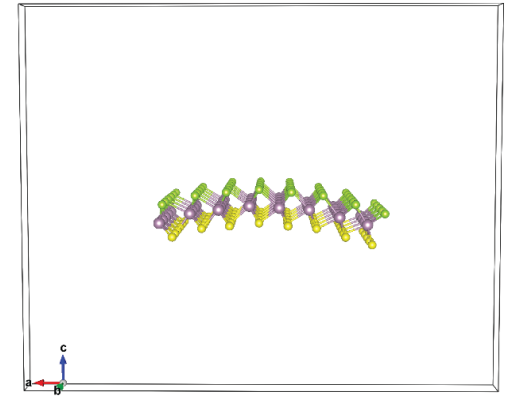

(c) Mo$^S_{Se}$ nanosheet curved towards Se    (f) Mo$^S_{Se}$ nanosheet curved towards Se

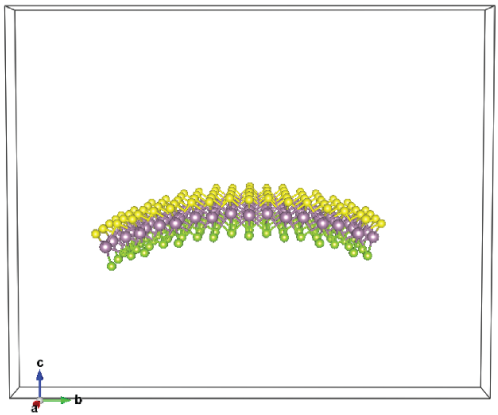 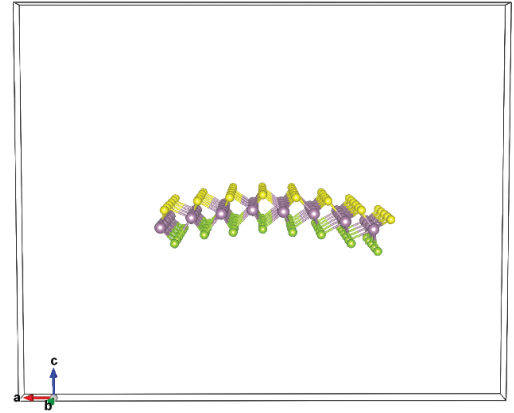

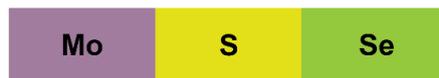

Mo    S    Se

**Figure S7**: **a.**, **b.** and **c.** show the flat, curved towards S face and curved towards Se face armchair nanoribbons of Mo$^s$$_{Se}$. **d.**, **e.** and **f.** show the same configurations for zigzag nanoribbons of Mo$^S$$_{Se}$. The radius of curvature in all curved sheets is 30 Å

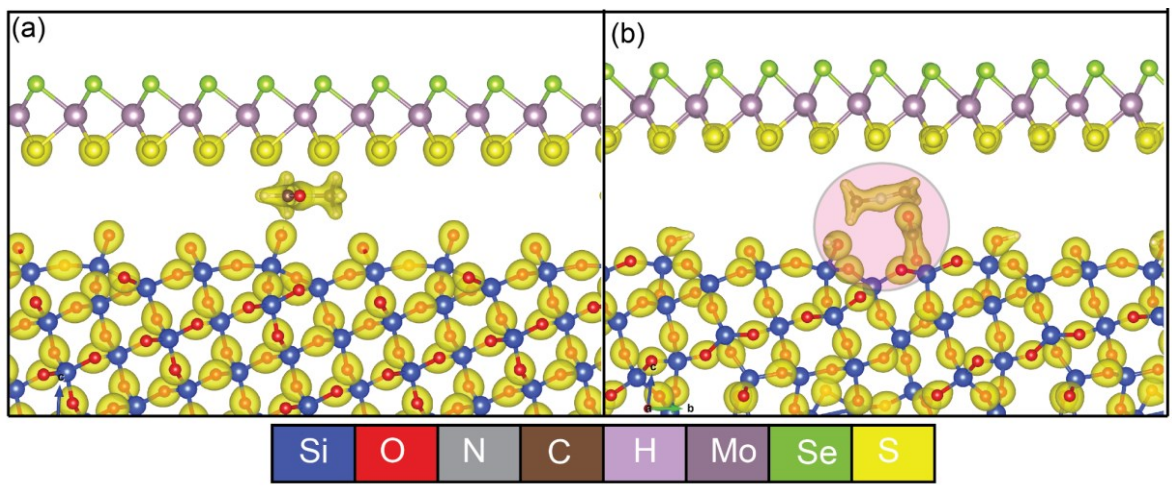

**Figure S8**: **a.** Shows the initial and **b.** shows the final timestep of the MD simulation with the charge densities plotted alongside the atoms. It can be seen that there is a hybridisation of the charge densities of the DMF molecule and the substrate, indicating strong bonding. All charge densities are plotted using an isosurface level of 0.14 $e$/Å$^3$

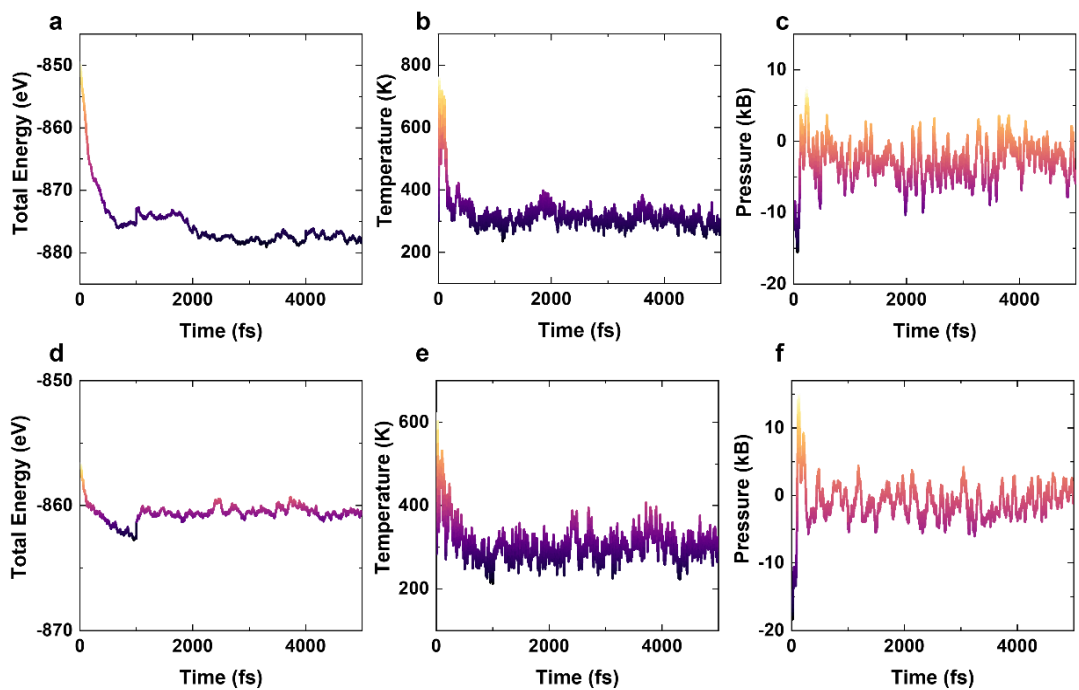

**Figure S9 a.** Total energy, **b.** temperature and **c.** total Pressure for intercalated DMF, and **d.** total energy, **e.** temperature and **f.** total Pressure for non-DMF intercalated heterostructure of $MoS^{Se}_S$ and SiO$_2$ substrate.

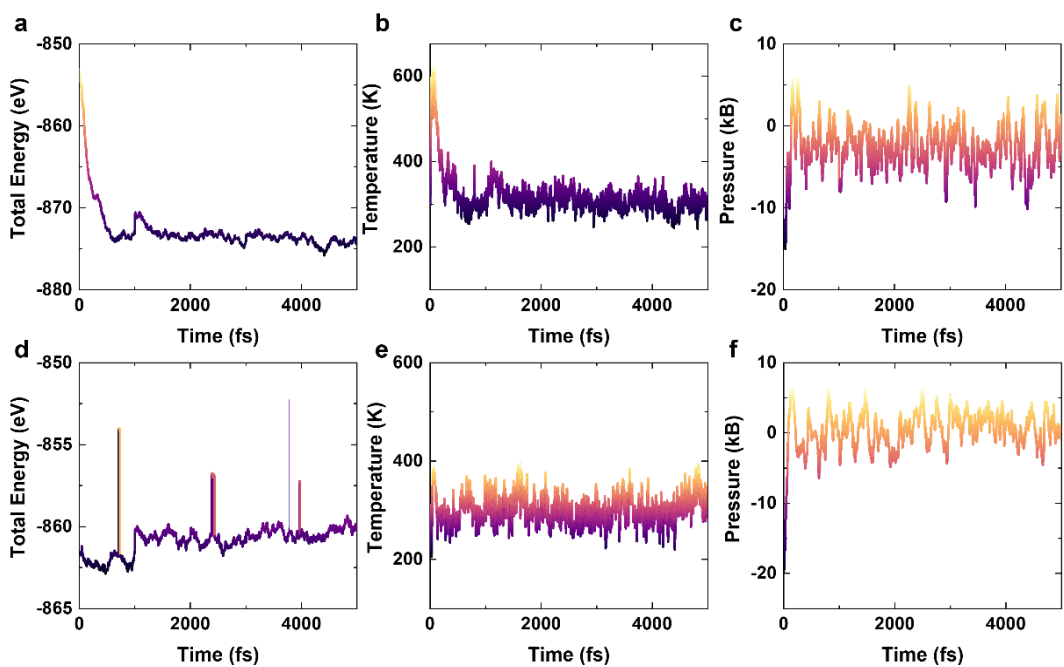

**Figure S10 a.** Total energy, **b.** temperature and **c.** total pressure for intercalated DMF, and **d.** total energy, **e.** temperature and **f.** total pressure for non-DMF intercalated heterostructure of $Mo_{Se}^{S}$ and SiO2 Substrate

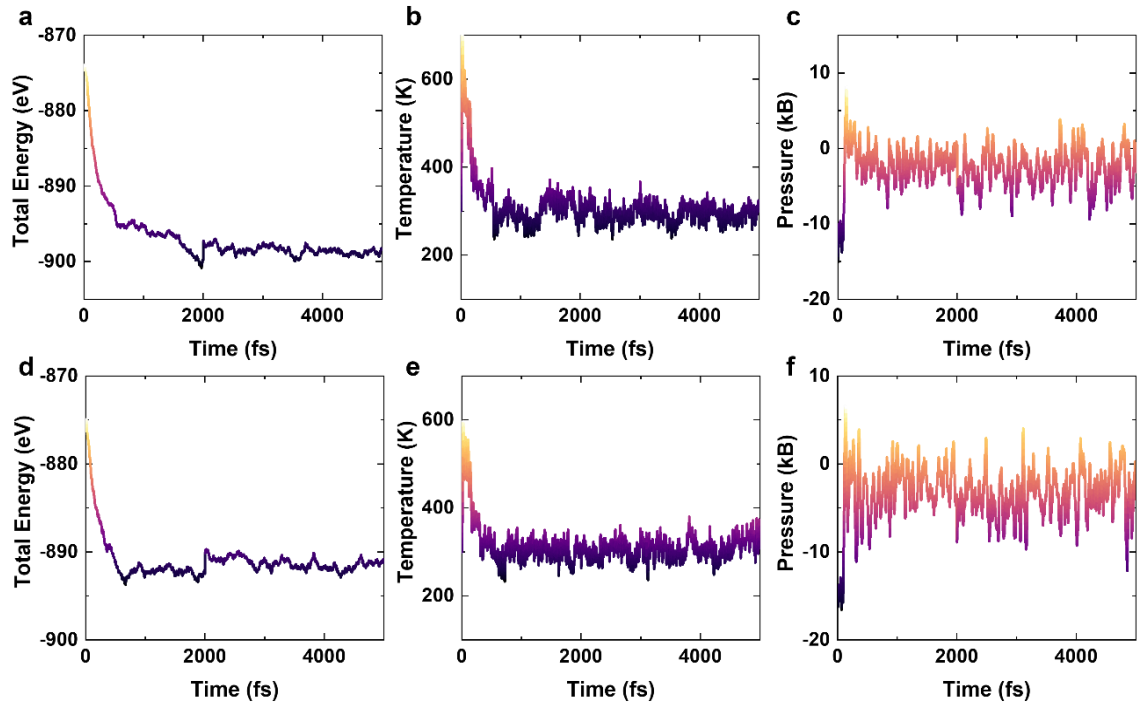

**Figure S11 a.** Total energy, **b.** temperature and **c.** total pressure for intercalated DMF, and **d.** total energy, **e.** temperature and **f.** total pressure for non-DMF intercalated heterostructure of $W_{S}^{Se}$ and SiO₂ Substrate

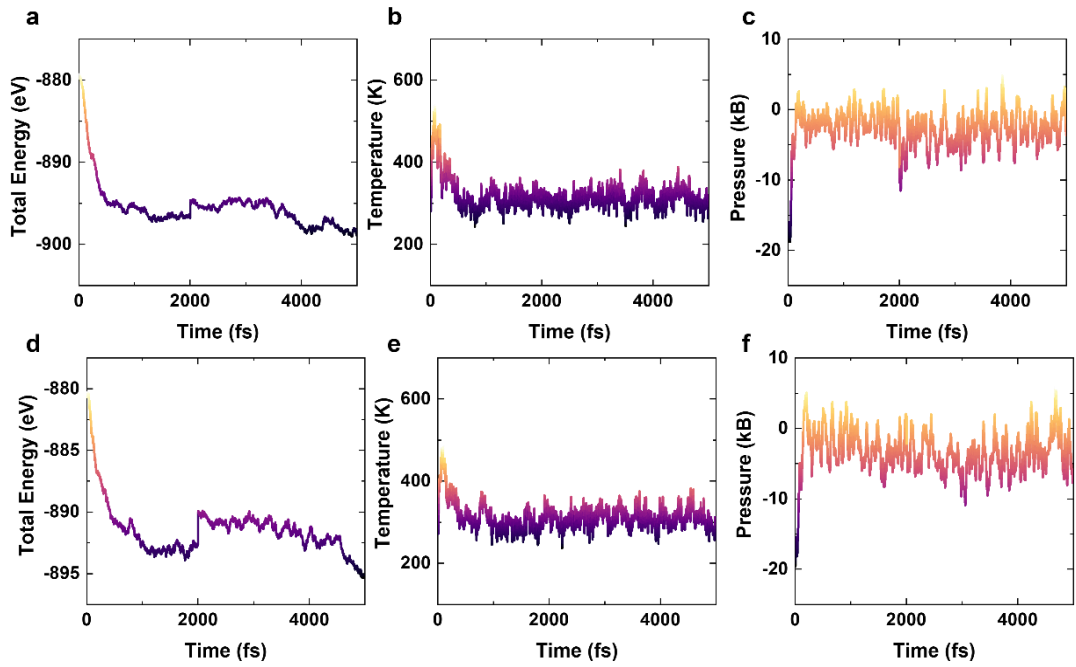

**Figure S12 a.** Total energy, **b.** temperature and **c.** total pressure for intercalated DMF, and **d.** total energy, **e.** temperature and **f.** total pressure for non-DMF intercalated heterostructure of $W^S_{Se}$ and SiO$_2$ Substrate

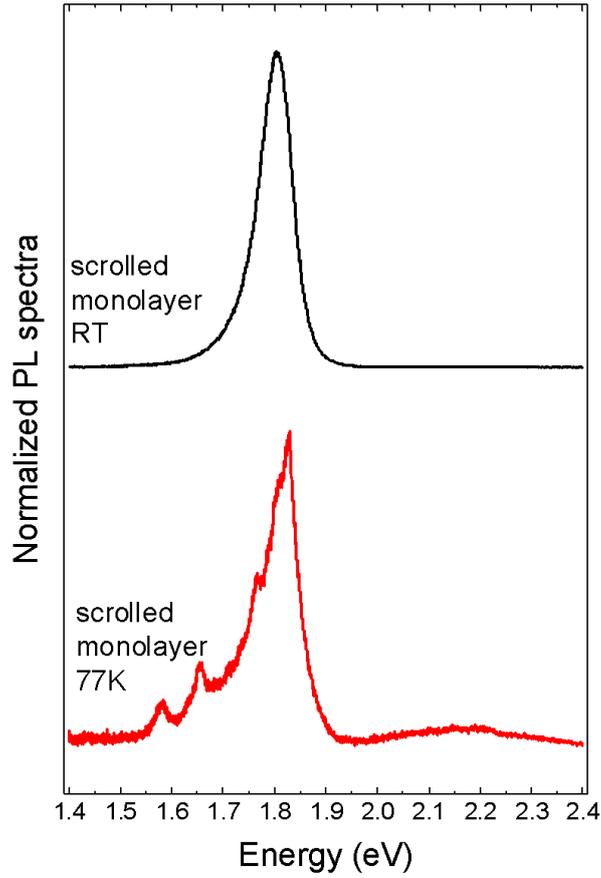

**Figure S13.** Comparison of Photoluminescence Spectra of Scrolled Monolayer $W_{Se}^{S}$ At 77K, Room-T was Created from tungsten selenide as the starting material.

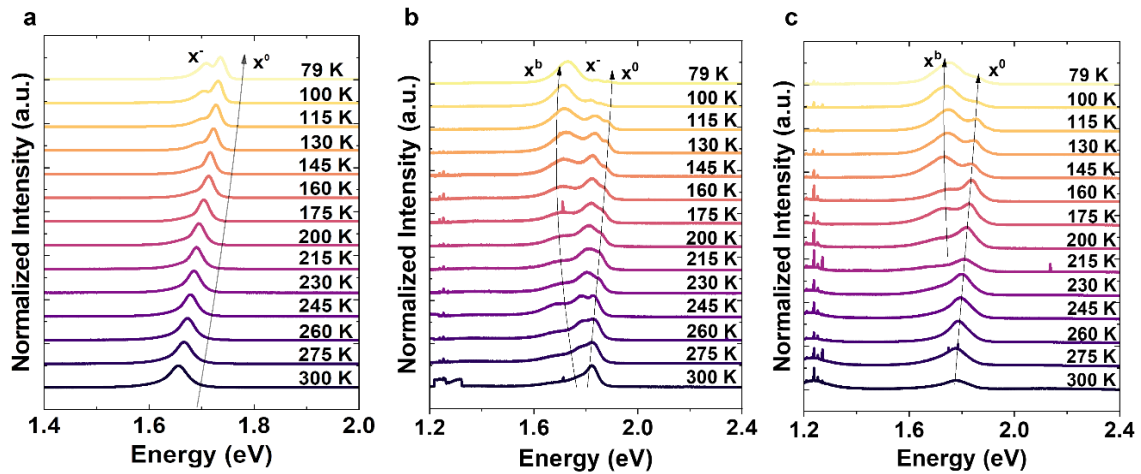

**Figure S14 a.** Temperature-Dependent Photoluminescence Spectra of pristine Monolayer $WSe_2$ **b.** Temperature Dependent Photoluminescence Spectra of defected Monolayer $W_{Se}^{S}$ flat sample **c.** Temperature Dependent Photoluminescence Spectra of defected Monolayer $W_{Se}^{S}$ scrolled sample

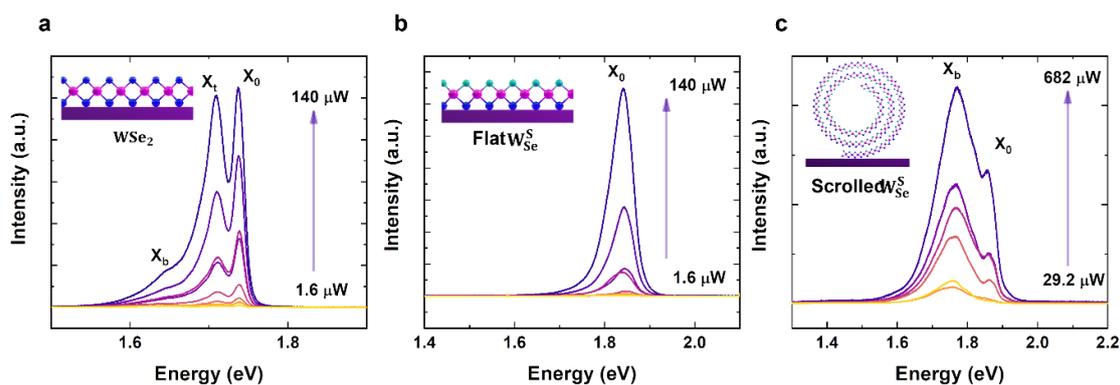

**Figure S15 a.** Power Dependent Photoluminescence Spectra of Monolayer WSe$_2$ **b.** Power Dependent Photoluminescence Spectra of Monolayer $W^S_{Se}$ flat sample **c.** Power Dependent Photoluminescence Spectra of Monolayer $W^S_{Se}$ scrolled sample

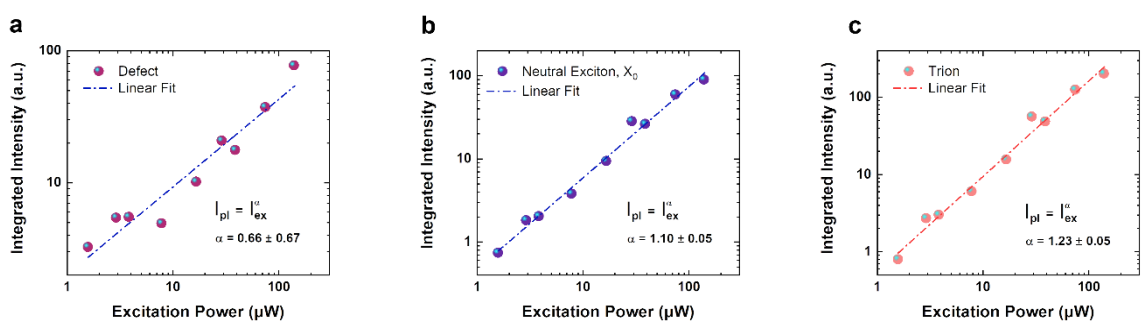

**Figure S16 a.** Integrated Peak Intensity vs excitation power measurement on monolayer WSe$_2$ (Bound exciton /X$^b$) **b.** Integrated Peak Intensity vs excitation power measurement on monolayer WSe$_2$ (Neutral exciton /X$^0$) **c.** Integrated Peak Intensity vs excitation power measurement on monolayer WSe$_2$ (Trion/X$^0$)

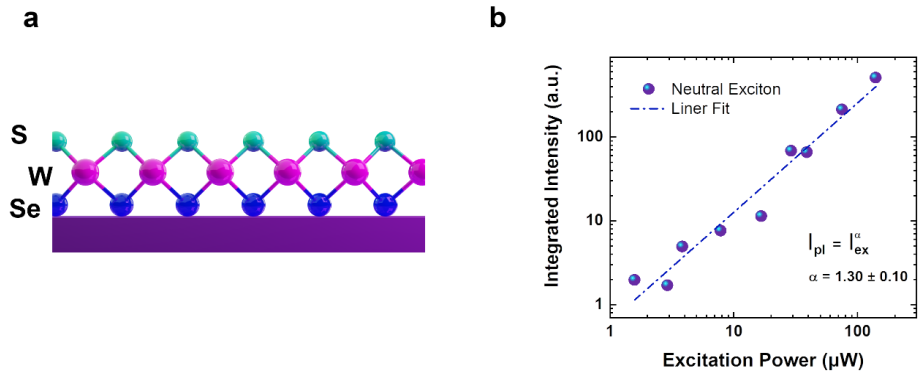

**Figure S17 a.** Configuration of $W^S_{Se}$ flat sample during the power-dependent PL measurement **b.** Integrated Peak Intensity vs excitation power measurement on monolayer $W^S_{Se}$ flat sample (Neutral exciton /$X^0$)

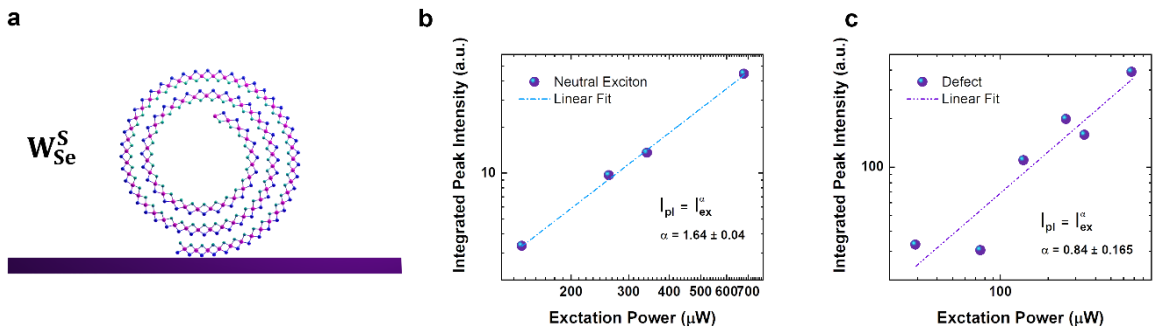

**Figure S18: -** Configuration of $W^S_{Se}$ scrolled sample during the power-dependent PL measurement **b.** Integrated Peak Intensity vs excitation power measurement on monolayer $W^S_{Se}$ scrolled sample (Neutral exciton /$X^0$) **c.** Integrated Peak Intensity vs excitation power measurement on monolayer $W^S_{Se}$ scrolled sample (bound exciton/$X^b$)

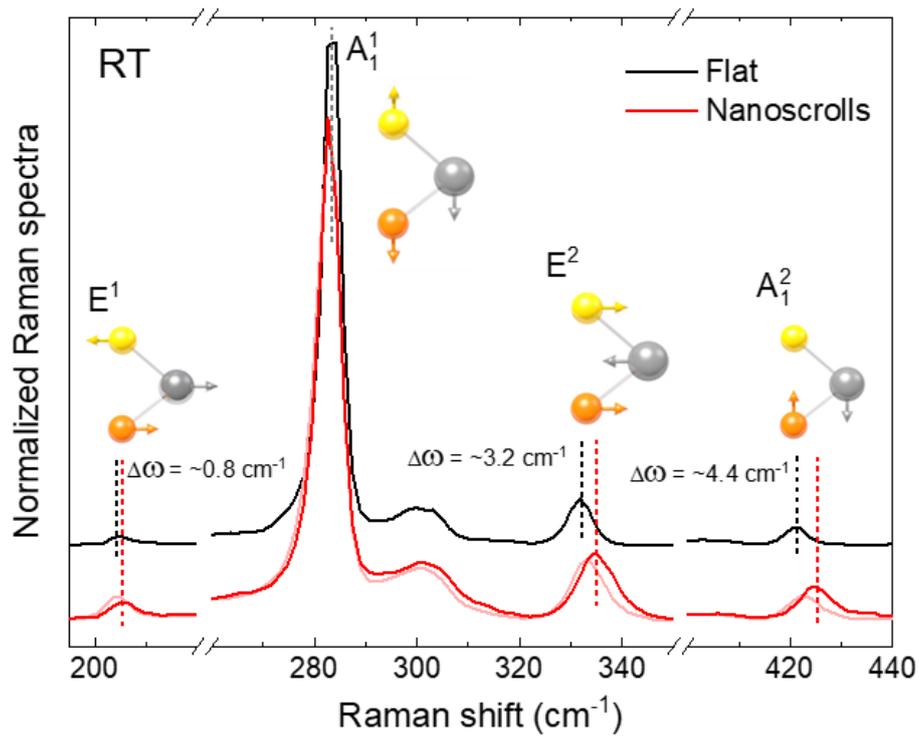

**Figure S19:** - Raman spectra, of 2D Janus $W^S_{Se}$ monolayer before and after scrolling at room temperature

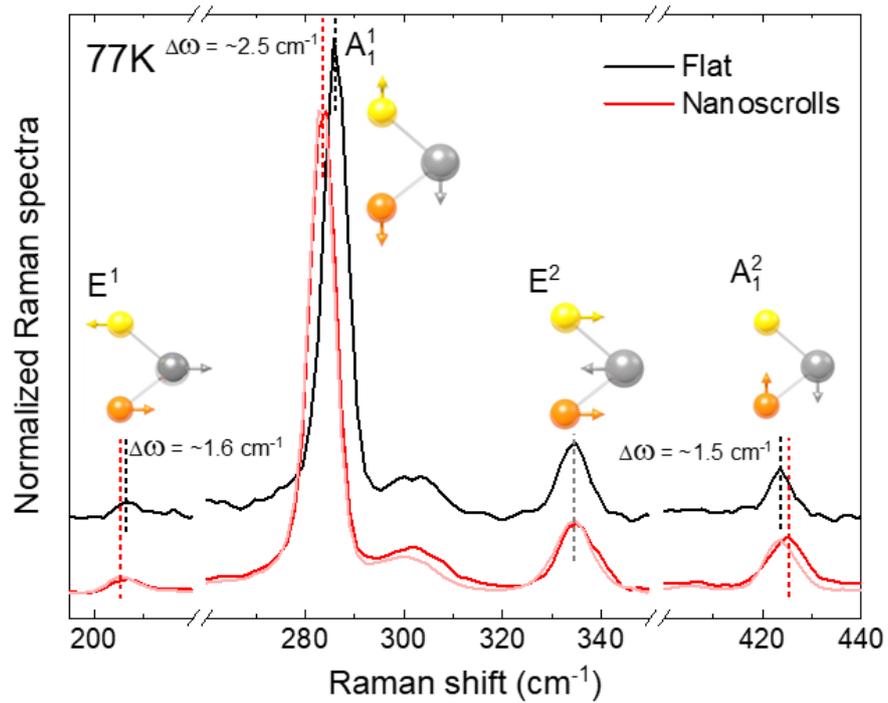

**Figure S20:** - Raman spectra, of 2D Janus $W_{Se}^{S}$ monolayer before and after scrolling at 77K

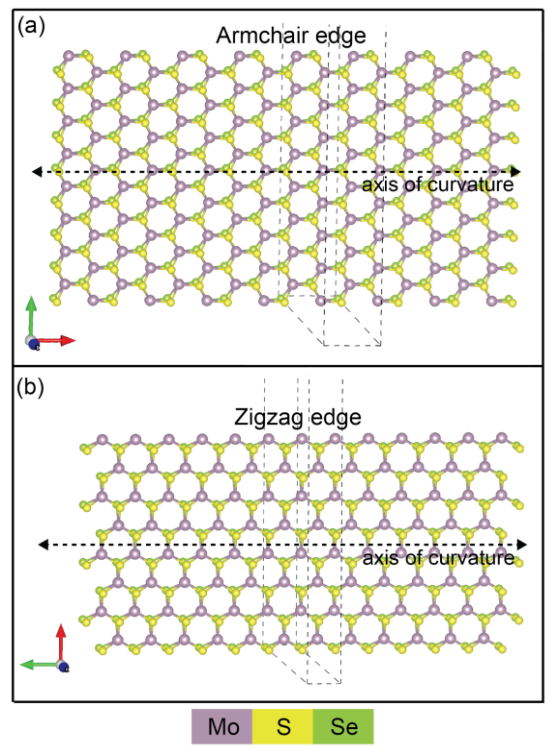

**Figure S21:** - **a.** *z*-projection of flat nanoribbon simulated with armchair edges and **b.**) *z*-projection of flat nanoribbon simulated with zigzag edges. Red, green and blue directions are *x*, *y* and *z*, respectively.

Nanoribbon is finite in the *x* direction and infinite in the *y* direction. The dotted axes in both figures indicate the axis along which the flat nanoribbon is curved to generate curved structures. Dotted box boundaries show the simulation cell.

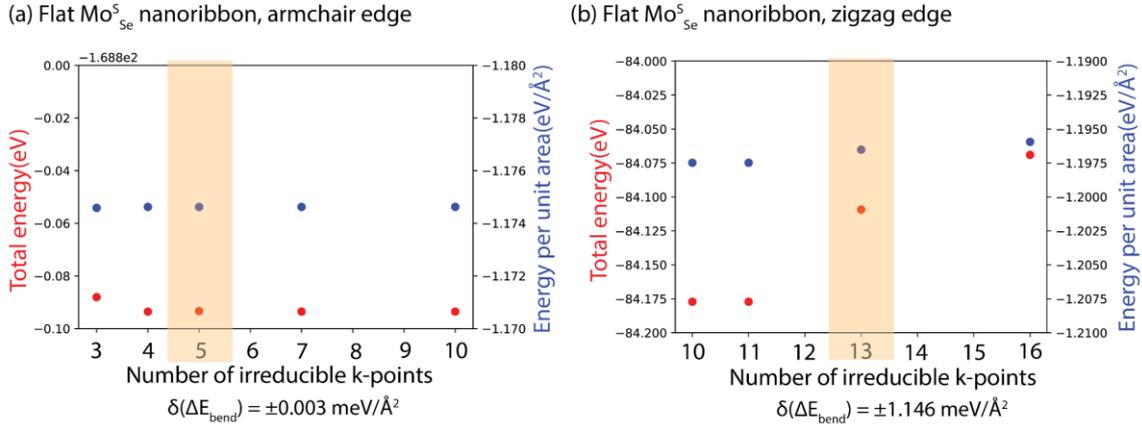

**Figure S22: - a.** and **b.** show the total Energy (red) and Energy per unit area (blue) for an armchair and zigzag flat $Mo^S_{Se}$ nanoribbon, respectively. The yellow highlighted rectangle indicates the *k*-points selected for all the simulations.

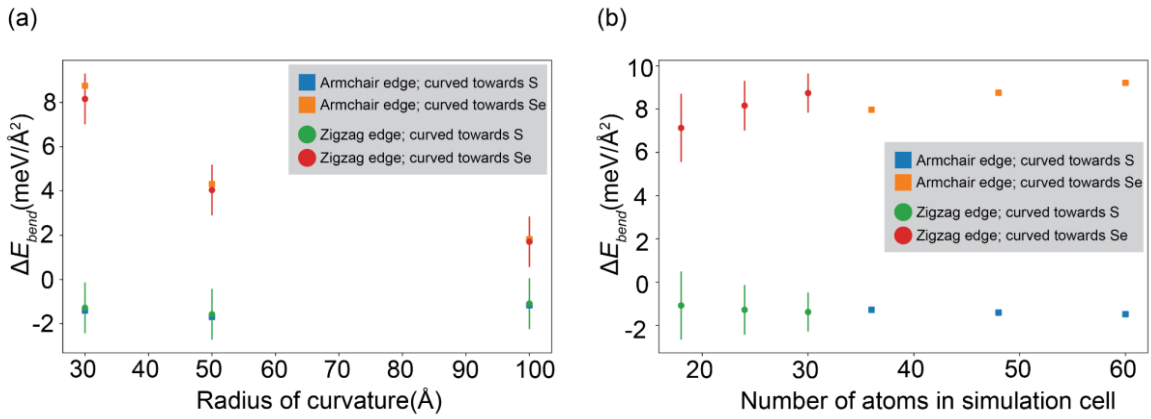

**Figure S23: - a.** shows the variation of $\Delta E_{bend}$ with the radius of curvature and **b.** shows the variation of $\Delta E_{bend}$ with the number of atoms in the simulation cell for both armchair and zigzag nanoribbons curved towards the S and Se faces. It can clearly be seen that in all cases bending towards the S face is more stable than bending towards the Se face.

| 2D Material | Presence of intercalated DMF | z-separation(Å) | E(eV) |
|---|---|---|---|
|  | Yes | 6.31 | -874 |
| $Mo_{Se}^{S}$ | No | 4.76 | -860 |
| $Mo_{S}^{Se}$ | Yes | 6.71 | -877 |
| | No | 4.29 | -860 |
| $W_{Se}^{S}$ | Yes | 6.45 | -898 |
| | No | 3.65 | -893 |
| $W_{S}^{Se}$ | Yes | 7.19 | -898 |
| | No | 4.45 | -891 |

**Table S1**: The average z-separation and total energy of configuration for the last 500-time steps of each MD simulated configuration. The aggregate parameters, such as the total energy, temperature, and total pressure of the system, are plotted as a function of the time step in Fig. S9-S12. These figures show that all the parameters converged within the 5 ps run time. Therefore, we consider the average values of the parameters for the last 500-time steps to be the equilibrium values of the parameters. In Table S1, we list the average values of the total energy to conclude the relative stability of the structures (see the previous section).

| Edge | Number of atoms in the simulation cell | Radius of curvature(Å) | Curved configuration | $\Delta E_{bend}$(meV/Å$^2$) |
|---|---|---|---|---|
| Armchair | 48 | 30 | Curved towards S face | -1.407±0.003 |
| | | | Curved towards Se face | 8.754±0.003 |
| Armchair | 48 | 50 | Curved towards S face | -1.691±0.003 |
| | | | Curved towards Se face | 4.313±0.003 |
| Armchair | 48 | 100 | Curved towards S face | -1.164±0.003 |
| | | | Curved towards Se face | 1.818±0.003 |

| Edge | Number of atoms in the simulation cell | Radius of curvature(Å) | Curved configuration | $\Delta E_{bend}$(meV/Å$^2$) |
|---|---|---|---|---|
| Zigzag | 24 | 30 | Curved towards S face | -1.281±1.146 |
| | | | Curved towards Se face | 8.161±1.146 |
| Zigzag | 24 | 50 | Curved towards S face | -1.571±1.146 |
| | | | Curved towards Se face | 4.043±1.146 |
| Zigzag | 24 | 100 | Curved towards S face | -1.090±1.146 |
| | | | Curved towards Se face | 1.708±1.146 |

**Table S2**: Values of ΔE$_{bend}$ for different radii of curvature for armchair and zigzag Mo$^S_{Se}$ nanoribbons.

| Edge | Number of atoms in the simulation cell | Radius of curvature(Å) | Curved configuration | $\Delta E_{bend}$(meV/Å$^2$) |
|---|---|---|---|---|
| Armchair | 36 | 30 | Curved towards S face | -1.279±0.003 |
| | | | Curved towards Se face | 7.970±0.003 |
| Armchair | 48 | 30 | Curved towards S face | -1.407±0.003 |
| | | | Curved towards Se face | 8.754±0.003 |
| Armchair | 60 | 30 | Curved towards S face | -1.482±0.003 |
| | | | Curved towards Se face | 9.212±0.003 |
| Zigzag | 18 | 30 | Curved towards S face | -1.081±1.146 |
| | | | Curved towards Se face | 7.129±1.146 |
| Zigzag | 24 | 30 | Curved towards S face | -1.281±1.146 |
| | | | Curved towards Se face | 8.161±1.146 |
| Zigzag | 30 | 30 | Curved towards S face | -1.383±1.146 |
| | | | Curved towards Se face | 8.740±1.146 |

**Table S3**: Values of $\Delta E_{bend}$ for different widths of the nanoribbons, expressed in terms of the number of atoms, for armchair and zigzag $Mo^S_{Se}$ nanoribbons.

**Section S2**

We use the Hetero2D package to generate the Janus-SiO$_2$ heterostructures for $M^X_Y$ (M=Mo/W; X, Y=S/Se). We apply constraints of 7% lattice mismatch between the original unit cells (h, k, l) with h, k, l <= 1 and a maximum interface surface area of 130 Å$^2$. We find that the (111) surface of SiO$_2$ is symmetry matched with the Janus materials. No other Janus-SiO$_2$ interfaces were found within this constraint.

We simulate a 4-layer (9.9 Å) slab of SiO$_2$ for each heterostructure. While the experiments are performed on an amorphous SiO$_2$ substrate, we use the crystalline $\beta$-cristobalite phase for our calculations. First, we simulate the Janus monolayers placed at an initial distance of 5.6 Å from the SiO$_2$ substrate. Then, in each heterostructure slab, we add a single DMF molecule between the Janus monolayer and the substrate. The DMF molecule is initially placed at an initial distance of 2.2. Å below the Janus monolayer.

We also simulate a heterostructure configuration where the DMF molecule is placed far away (17.6 Å) from the SiO$_2$ -Janus interfaces to conserve the total number of atoms in the unit cell while at the same time negating the interaction between the DMF molecule and the interface.

The confidence interval on the values of $z$-separation (plotted as shaded regions in Figure 3c, d-g, h) for every time step $t$ is given by

$$CI(d,t) = (d(t) - (z(t)_X^{max} - z(t)_{SiO_2}^{min}), d(t) - (z(t)_X^{min} - z(t)_{SiO_2}^{max}))$$

Where $d(t)$ is the $z$-separation at the t$^{th}$ time step $z(t)_X^{max/min}$ and $z(t)_{SiO_2}^{max/min}$ are the maximum(minimum) $z$-positions of the bottom chalcogenide layer atoms and top SiO$_2$ substrate layer atoms, respectively.

We determine the errors in the values of $\Delta E_{bend}$ by first establishing the errors in the total energy as shown in figure S20 using the formula below:

$$\delta(\Delta E_{bend}) = \frac{2 \times \delta(E_{total})}{Area}$$

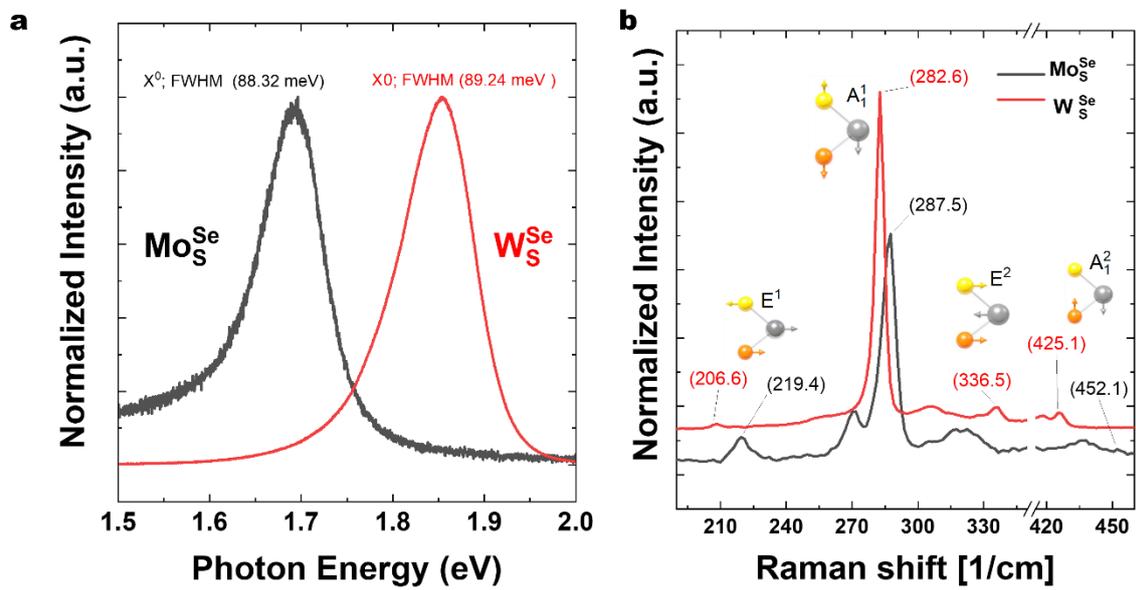

**Fig. S24**: **a.** Room Temperature Photoluminescence Spectra of Janus $W_S^{Se}$ and $Mo_S^{Se}$ synthesised from their exfoliated sulfur-based classical TMD. **b**. Raman Spectrum of 2D Janus $W_S^{Se}$ and $Mo_S^{Se}$ With their characteristic modes highlighted.

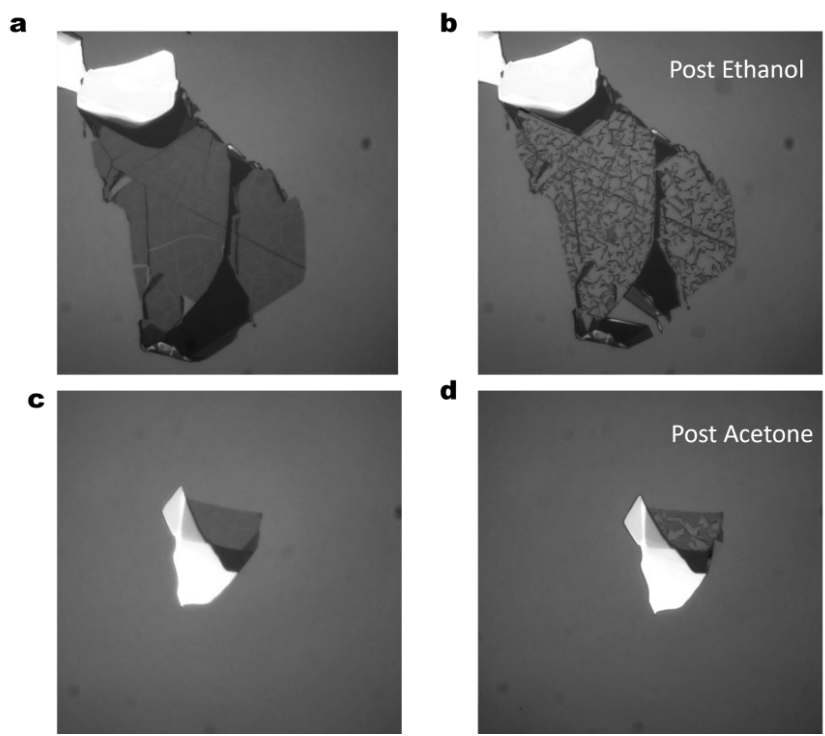

**Fig. S25: a, c** Monolayer Janus $W^S_{Se}$ synthesised from Exfoliated Monolayer WSe$_2$. **b.** Shows sample in panel **a** after treatment with 50 μL ethanol solution for < 5 seconds. **d.** Shows sample in panel **c** after treatment with 50 μL Acetone solution for < 5 Seconds

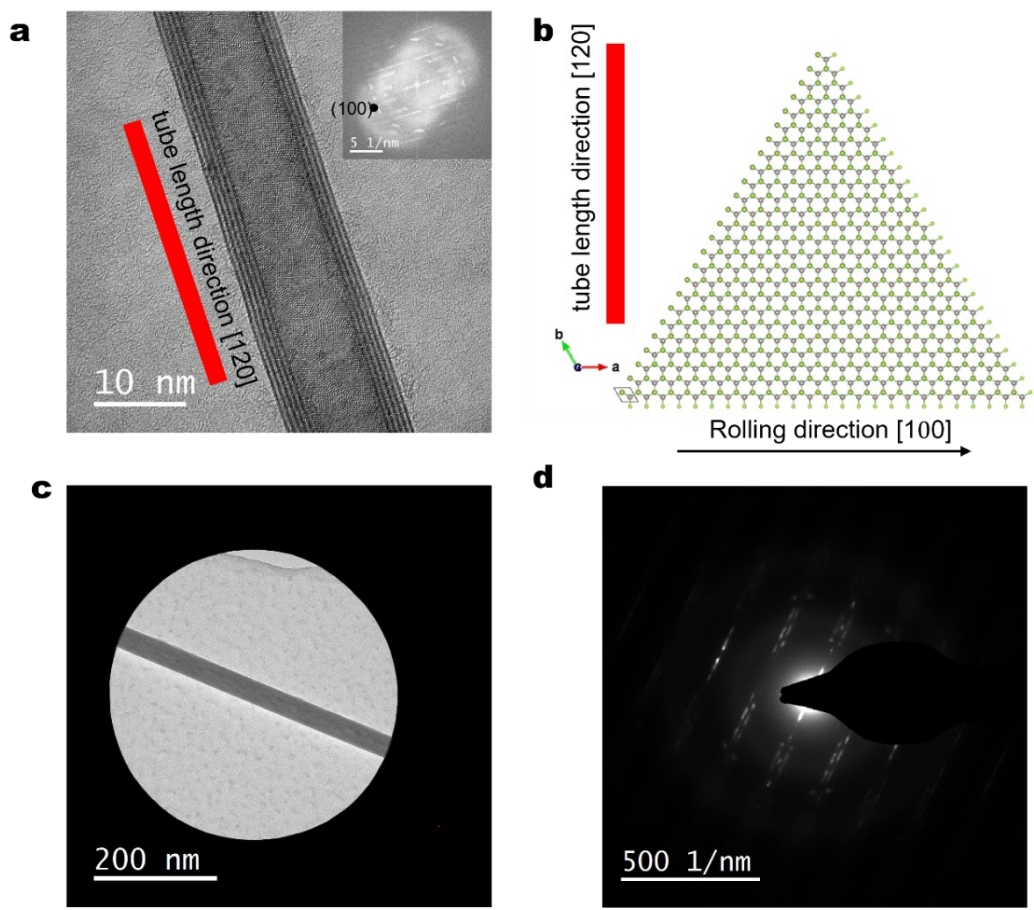

**Fig. S26 a.** HRTEM image of Janus $W^S_{Se}$, *inset:* FFT of the image is shown in the upper right, diffraction spot (100) is denoted by the dark spot. **b.** Schematic representation of rolling direction with respect to the tube length. **c.** The region is selected using a selective area aperture, and **d.** the corresponding diffraction pattern is shown on the right.

HRTEM images were taken on a Titan microscope at 80 kV, FFT of the image is shown in the upper right, diffraction spot (100) is denoted by the dark spot. This result reveals the rolling along [100] direction as shown in the scheme to the right. Similar results can also be recorded using electron diffraction mode. Here the TEM image shows the region selected using a selective area aperture, and the corresponding diffraction pattern is shown on the right. The diffraction also shows the rolling along [100] direction, and the tube length is along [120] direction.